\documentclass[]{spie}  %>>> use for US letter paper
\usepackage[]{graphicx}
\usepackage{float}% http://ctan.org/pkg/float

\title{The near infrared camera for the Subaru Prime Focus Spectrograph} 

\author{
Stephen A. Smee\supit{a},
James E. Gunn\supit{b},
Mirek Golebiowski\supit{a},
Robert Barkhouser\supit{a},
Sebastien Viv\`es\supit{c},
Sandrine Pascal\supit{c},
Michael Carr\supit{b},
Stephen C. Hope\supit{a},
Craig Loomis\supit{b},
Murdock Hart\supit{a},
Hajime Sugai\supit{d},
Naoyuki Tamura\supit{d},
and Atsushi Shimono\supit{d}
\skiplinehalf
\supit{a}Johns Hopkins Unversity, Department of Physics and Astronomy, 3701 San Martin Drive, Baltimore, MD 21218, USA; \\
\supit{b}Princeton University, Department of Astrophysical Sciences, Princeton, NJ 08544, USA; \\
\supit{c}Aix Marseille Universit\'e -CNRS, LAM (Laboratoire d'Astrophysique de Marseille), UMR 7326, 13388, Marseille, France; \\
\supit{d}Kavli Institute for the Physics and Mathematics of the Universe (WPI), The University of Tokyo, 5-1-5, Kashiwanoha, Kashiwa 277-8583, Japan; 
}

\authorinfo{Send correspondence to: smee@jhu.edu}
\begin{document} 
  \maketitle 

%%%%%%%%%%%%%%%%%%%%%%%%%%%%%%%%%%%%%%%%%%%%%%%%%%%%%%%%%%%%% 
\begin{abstract}
We present the detailed design of the near infrared camera for the SuMIRe (Subaru Measurement of Images and Redshifts) Prime Focus Spectrograph (PFS) being developed for the Subaru Telescope.  The PFS spectrograph is designed to collect spectra from 2394 objects simultaneously, covering wavelengths that extend from 380~nm -- 1.26~$\mu$m.  The spectrograph is comprised of four identical spectrograph modules, with each module collecting roughly 600 spectra from a robotic fiber positioner at the telescope prime focus.  Each spectrograph module will have two visible channels covering wavelength ranges 380~nm -- 640~nm and 640~nm -- 955~nm, and one near infrared (NIR) channel with a wavelength range 955~nm -- 1.26~$\mu$m.  Dispersed light in each channel is imaged by a 300~mm focal length, f/1.07, vacuum Schmidt camera onto a 4k x 4k, 15~$\mu$m pixel, detector format.  For the NIR channel a HgCdTe substrate-removed Teledyne 1.7~$\mu$m cutoff device is used.  In the visible channels, CCDs from Hamamatsu are used.  These cameras are large, having a clear aperture of 300~mm at the entrance window, and a mass of $\sim$ 250~kg.

Like the two visible channel cameras, the NIR camera contains just four optical elements: a two-element refractive corrector, a Mangin mirror, and a field flattening lens.  This simple design produces very good imaging performance considering the wide field and wavelength range, and it does so in large part due to the use of a Mangin mirror (a lens with a reflecting rear surface) for the Schmidt primary.  In the case of the NIR camera, the rear reflecting surface is a dichroic, which reflects in-band wavelengths and transmits wavelengths beyond 1.26~$\mu$m.  This, combined with a thermal rejection filter coating on the rear surface of the second corrector element, greatly reduces the out-of-band thermal radiation that reaches the detector. 

The camera optics and detector are packaged in a cryostat and cooled by two Stirling cycle cryocoolers.  The first corrector element serves as the vacuum window, while the second element is thermally isolated and floats cold.  An assembly constructed primarily of silicon carbide is used to mount the Mangin mirror, and to support the detector and field flattener.  Thermal isolation between the cold optics and warm ambient surroundings is provided by G10 supports, multi-layer insulation, and the vacuum space within the cryostat.  

In this paper we describe the detailed design of the PFS NIR camera and discuss its predicted optical, thermal, and mechanical performance.

\end{abstract}

%>>>> Include a list of keywords after the abstract 

\keywords{Near infrared camera, Spectrograph, Multi-object spectrograph, Subaru Telescope, SuMIRe, HgCdTe, Optomechanics}

%%%%%%%%%%%%%%%%%%%%%%%%%%%%%%%%%%%%%%%%%%%%%%%%%%%%%%%%%%%%%
\section{INTRODUCTION}
\label{sec:INTRODUCTION}

The SuMIRe Prime Focus Spectrograph~\cite{2012SPIE.8446E..0YS} (PFS) will consist of 4 identical spectrograph modules each receiving roughly 600 fibers from the Cobra prime-focus fiber placement robot.  Each spectrograph module will have three channels, covering the wavelength ranges 380~nm -- 640~nm, 640~nm -- 955~nm, and 955~nm -- 1.26~$\mu$m. The cameras for each channel are almost identical vacuum Schmidts.  Spectra are imaged onto detectors with a 4k x 4k format and 15 $\mu$m pixels.  A pair of 2k x 4k Hamamatsu CCDs comprise the detector array for the visible cameras~\cite{Pascal2014}.  For the near infrared (NIR) camera, which is the subject of this paper, the 1.7~$\mu$m cutoff Teledyne H4RG-15 Mercury-Cadmium-Telluride device is used.  The H4RG is described in a recent paper by Blank~\cite{2012SPIE.8453E..0VB}.

Since the near infrared camera does not go very far into the infrared, one can (with care) use a near-room-temperature spectrograph, but must be very careful of the in-beam thermal background and design the camera/dewar for the IR detector so as to minimize the diffuse background.  As such, the PFS spectrograph modules will be housed in an insulated and cooled spectrograph room located on the IR4 floor of the Subaru dome.  The room will be temperature controlled to 5~C +/-~1~C.

The spectrograph is a bench mounted design, hence all components are fixed relative to gravity, which reduces the design complexity as compared to Cassegrain and Nasmyth mounted instruments.

Subaru is a very big telescope (8.2~m), and to use 15~$\mu$m pixel devices efficiently (with reasonable sampling) one must use a very fast camera.  One arcsecond at f/1.0 projects to 40~$\mu$m on the focal plane.  Therefore we have settled on an f/1.07, 300~mm focal length, camera design.  The fibers are 128~$\mu$m in diameter, and with input microlenses to reduce the numerical aperture to a level satisfactory for the fibers (f/2.8 from the input f/2.2) subtend 1.1 arcseconds on the sky.  The fibers image to a circular spot about 54~$\mu$m diameter on the detector, or about 3.5 pixels.  It is not possible with reasonable designs to make fashionable refracting systems this fast, hence modified Schmidts are the only reasonable choice.  In a fast Schmidt, the detector must be in the beam, and so obscures light, so it is desirable to have the beam as big as possible and the detector package as small as possible, preferably no larger than the detector itself.  This is most easily accomplished by making the cryostat the camera housing, so the cold detector and all the reflective optics are in vacuum.  The Schmidt design lends itself to this, since there is a corrector plate which can be made thick which can serve as a vacuum window.  This design also can accommodate a cold thermal shield, which turns out to be necessary for good performance in the near infrared.  The beam size in this design is set by the maximum size of VPH gratings currently available from Kaiser Optical Systems.

%%%%%%%%%%%%%%%%%%%%%%%%%%%%%%%%%%%%%%%%%%%%%%%%%%%%%%%%%%%%%
\section{DESIGN OVERVIEW}
\label{sec:DESIGN_OVERVIEW} 

\begin{figure}
   \begin{center}
   \begin{tabular}{c}
   \includegraphics[scale=0.35]{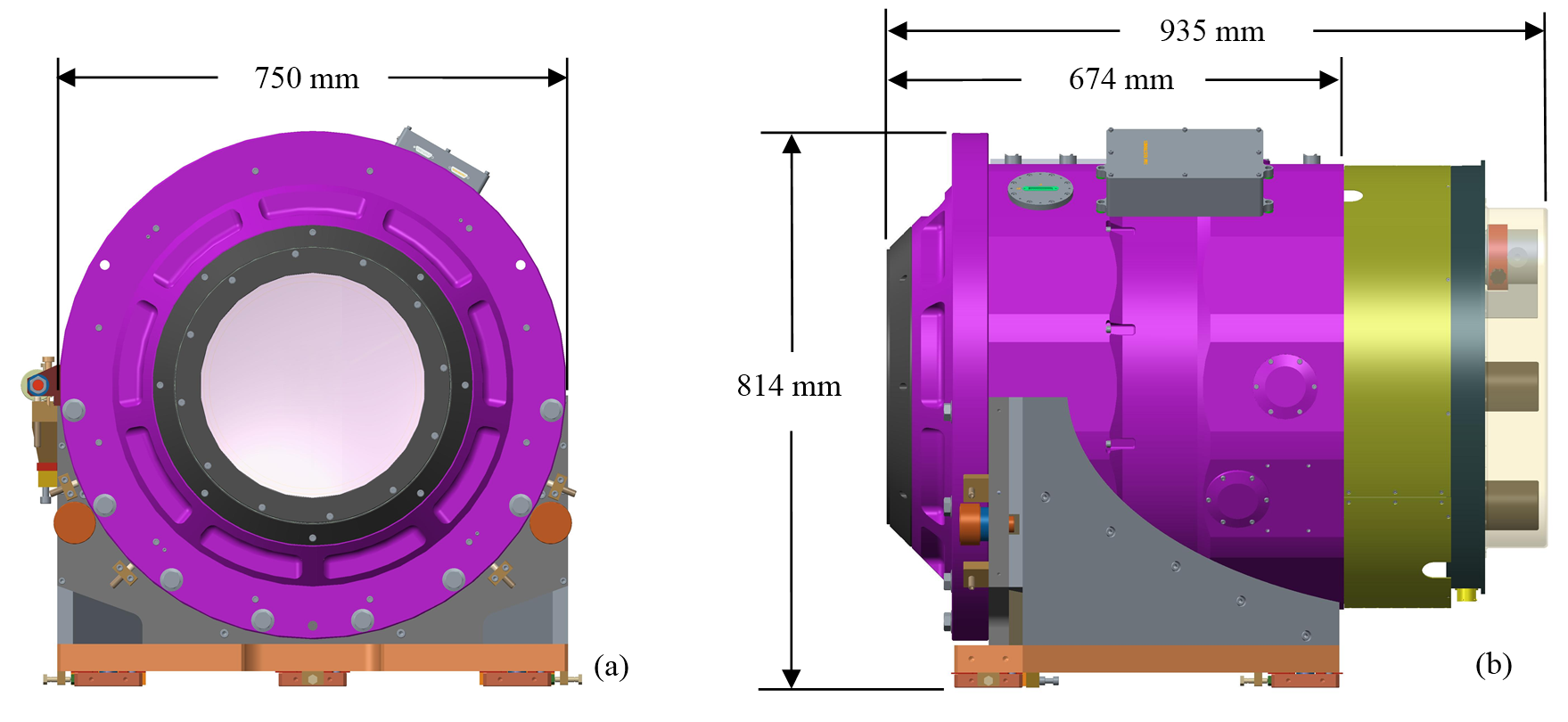}
   \end{tabular}
   \end{center}
   \caption[example] 
%>>>> use \label inside caption to get Fig. number with \ref{}
   { \label{fig:Fig1} 
Front (a) and side (b) views of the PFS near infrared camera supported in its mount.  The top of the camera sits 814~mm from the surface of the optical bench. The diameter across the front flange is 750~mm.  The overall length of the camera is 935~mm and the length of the cryostat is 674~mm.}
   \end{figure}

Figure \ref{fig:Fig1} shows the front and side of the near infrared camera supported in its cradle mount.  The base of the mount interfaces to the spectrograph module optical bench.  From the base, the camera sits 814~mm tall, has an overall length of 935~mm and is just over 750~mm wide when the roll adjustment mechanism is factored in.  The mass of the entire assembly is approximately 320~kg.  The mount itself has a mass of 70~kg.

%-------------
%-------------

The cradle mount design, by Winlight Systems\footnote{Winlight Systems, www.winlight-system.com}, is common to all cameras.  It has a three point kinematic interface to the optical bench at the base of the mount.  Lateral position and orientation of the camera are adjusted using adjustment screws at each of the three mounting pads.  Shims between the cradle base and each mounting pad allow for tip and tilt adjustment.  Roll adjustment to orient the detector to the spectra is made possible by a simple push/pull mechanism that rolls the entire cryostat relative to the mount.  Once adjusted, the location of the mounting pads are locked and the camera and cradle can be removed as an assembly and replaced accurately without need for further adjustment.

\begin{figure}[H]
   \begin{center}
   \begin{tabular}{c}
   \includegraphics[scale=0.62]{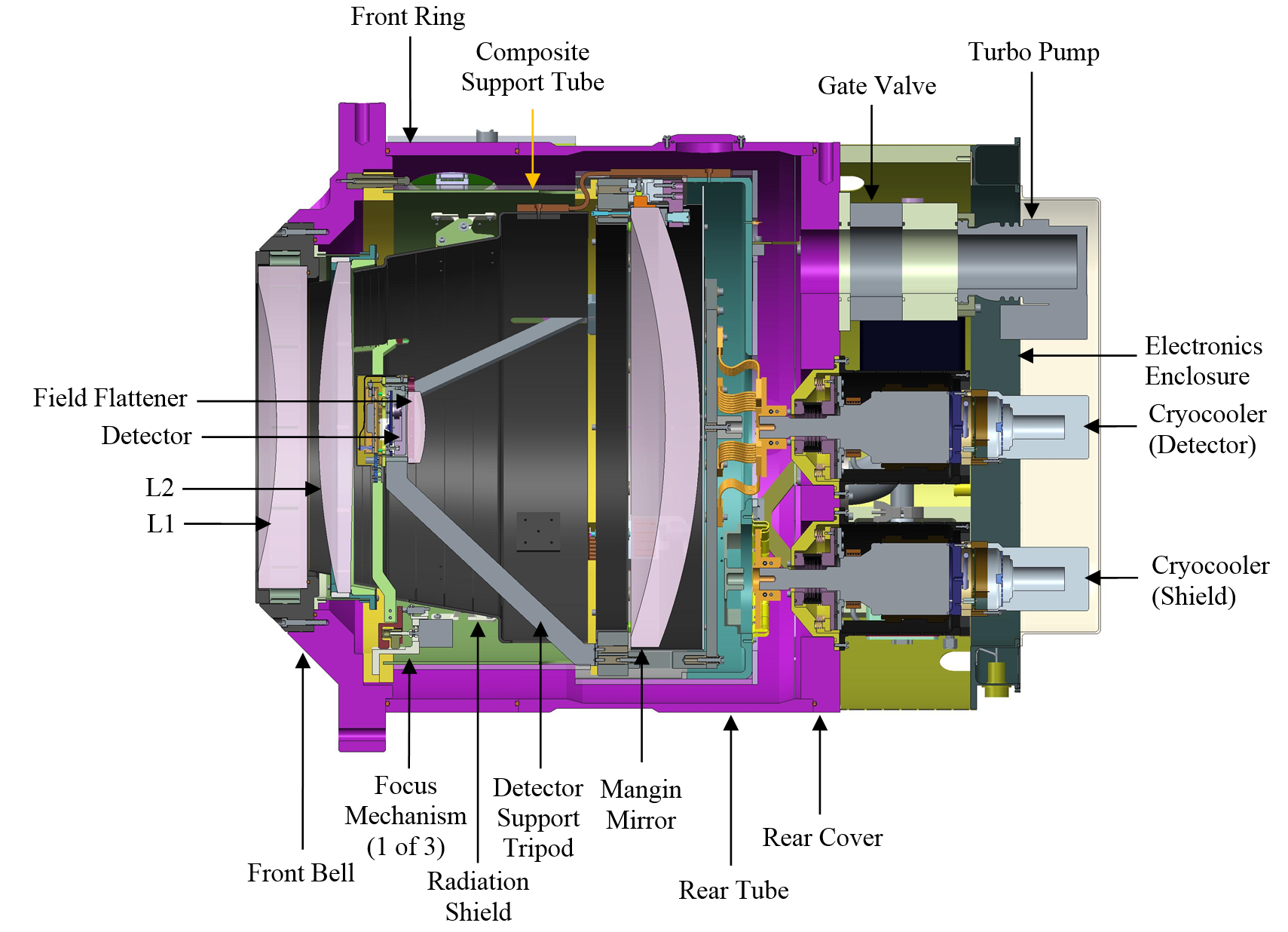}
   \end{tabular}
   \end{center}
   \caption[example] 
%>>>> use \label inside caption to get Fig. number with \ref{}
   { \label{fig:Fig2} 
Cross-section of the near infrared camera.  The camera contains four optical elements, a dual element corrector, a Mangin mirror, and a field flattener.  The detector resides at the focus of the mirror and is supported by a silicon carbide tripod.  The optics are housed within a cryostat constructed from four main structural components: the Front Bell, the Front Ring, the Rear Tube, and Rear Cover.  Mounted on the rear cover of the cryostat are two cryocoolers, two ion pumps, a turbo pump, and a vacuum gauge.  Electronics to control and monitor these components, as well as to monitor temperatures throughout the cryostat, are housed in a glycol-cooled electronics enclosure, which is also located at the rear of the cryostat.}
   \end{figure}

Figure \ref{fig:Fig2} shows a cross-section of the camera.  The camera contains four optical elements, a two-element corrector, a Mangin mirror, and a field flattening lens.  The detector resides at the focus of the mirror and is supported by a silicon carbide tripod.  The optics are housed within a cryostat constructed from four main structural components: the Front Bell, the Front Ring, the Rear Tube, and Rear Cover.  A cooled light baffle, i.e. a radiation shield, between the mirror and corrector reduces stray light and stray emission within the detector bandpass.  Mounted on the rear cover of the cryostat are two cryocoolers, two ion pumps, a turbo pump, and a vacuum gauge.  Electronics to control and monitor these components, as well as to monitor temperatures throughout the cryostat, are housed in a glycol-cooled electronics enclosure, which is also located at the rear of the cryostat.  

The cryostat body is 7075-T6 aluminum.  The use of the high strength aluminum alloy is prefered by the fabricator; they feel it is more stable than the more common 6061-T6 alloy.  Viton O-rings seal the cryostat interfaces and will be baked to ensure low permeation and outgassing.   The front bell, at the front of the cryostat, is the primary optomechanical interface.  As discussed below, the large on-axis bore establishes the optical axis and locates the corrector assembly; the remaining optics are aligned to this axis.  There are numerous access ports: three for connecting thermal straps that bridge the radiation shield to the thermal pan in the rear; and two to access fasteners connecting the cyrocooler tips to their respective thermal straps.  All electrical signals pass through hermetic feedthroughs located on the front ring.  The structural integrity of the cryostat was analyzed by finite element analysis and found to be very stiff.  Pressure loading produces a 20~$\mu$m translation of the corrector, and a $\sim$ 0.3~mm displacement of the rear cover.

Subaru does not allow liquid nitrogen for cooling instruments.  Detectors must be coooled using closed cycle crycoolers.  As discussed in Section~\ref{sec:THERMAL_DESIGN} we have chosen to use Sunpower Cryotel GT\footnote{Sunpower Inc., www.sunpowerinc.com} cryocoolers in this design; two of them.  One for the primary purpose of cooling the detector, and a second for the primary purpose of cooling the radiation shield.  There are a number of advantages to the Cryotel coolers, however they have one drawback: they vibrate.  Two moving pistons inside the unit oscillate at a frequency of 60~Hz, hence the coolers vibrate at a fundamental frequency of 60~Hz, with many higher order harmonics.  With the passive damper, Sunpower advertises an on-axis vibration level of approximately 400~mG at 60~Hz.  Measurements in our lab confirm this number; we measured 436~mG.  Those measurements also revealed comparable levels, 380~mG, off-axis but at 120~Hz.  Two strategies have been employed to mitigate the problem: first, develop an active damping system in concert with Sunpower (see Hope et al.\cite{Hope2014}, this conference); and second, design the camera to be {\it very} stiff, with all strucutural components in the load path from the cooler to the detector having a fundamental frequency higher than 120~Hz.  It is important to note that vibration amplitudes scale inversly as the square of the frequency, hence the higher order harmonics generated within the cooler have little, if any, impact on the optomechanical stability of the camera; we mainly care about the first two modes, 60~Hz and 120~Hz.  

%-------------
%-------------

%%%%%%%%%%%%%%%%%%%%%%%%%%%%%%%%%%%%%%%%%%%%%%%%%%%%%%%%%%%%%
\section{OPTICAL DESIGN}
\label{sec:OPTICAL_DESIGN} 

\begin{figure}
   \begin{center}
   \begin{tabular}{c}
   \includegraphics[scale=0.45]{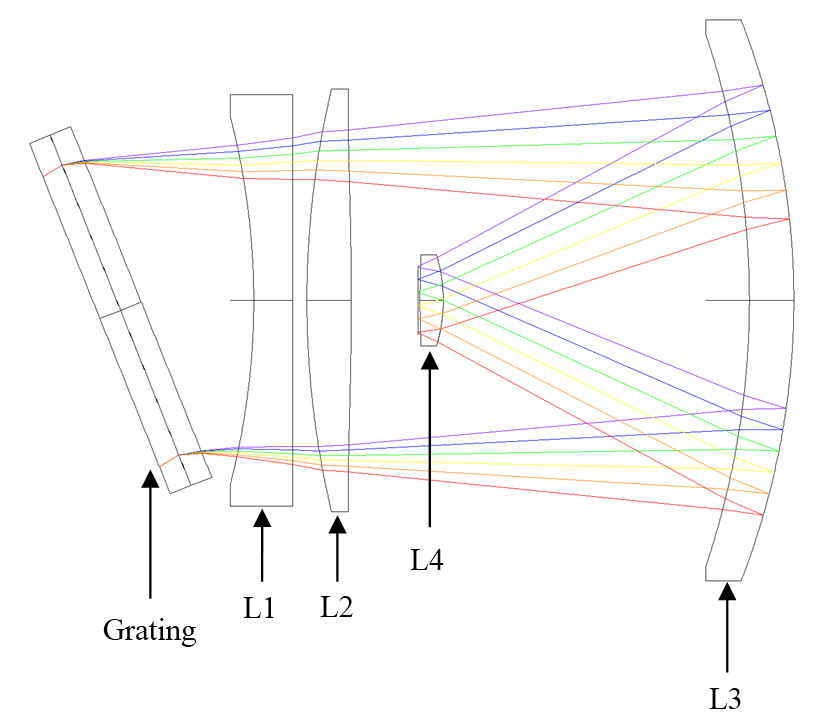}
   \end{tabular}
   \end{center}
   \caption[example] 
%>>>> use \label inside caption to get Fig. number with \ref{}
   { \label{fig:Fig3} 
Schematic depicting the near infrared camera optical layout.  The grating, shown to the left, is included for reference.  Light enters from the left, is transmitted through a dual-element corrector (L1 \& L2), is reflected from the rear surface of a Mangin mirror (L3), and comes to focus on the detector which resides just behind the field flattener (L4).  All four camera elements are fused silica.}
   \end{figure}

As is the case with the two visible channel cameras\cite{Pascal2014}, the NIR camera contains just four optical elements: a two-element refractive corrector, a Mangin mirror, and a field flattening lens; all of which are fused silica; see Figure \ref{fig:Fig3}.  The optical prescription is provided in Table~\ref{tab:Tab1}.  This simple design produces very good imaging performance considering the wide field and wavelength range, and it does so in large part due to the use of a Mangin mirror (using the term in a general sense to refer to a lens with a reflecting rear surface) for the Schmidt primary. 

\begin{table*}
\begin{center}
\caption{Optical Prescription for the near infrared camera\label{tab:Tab1}}
\includegraphics[scale=0.42]{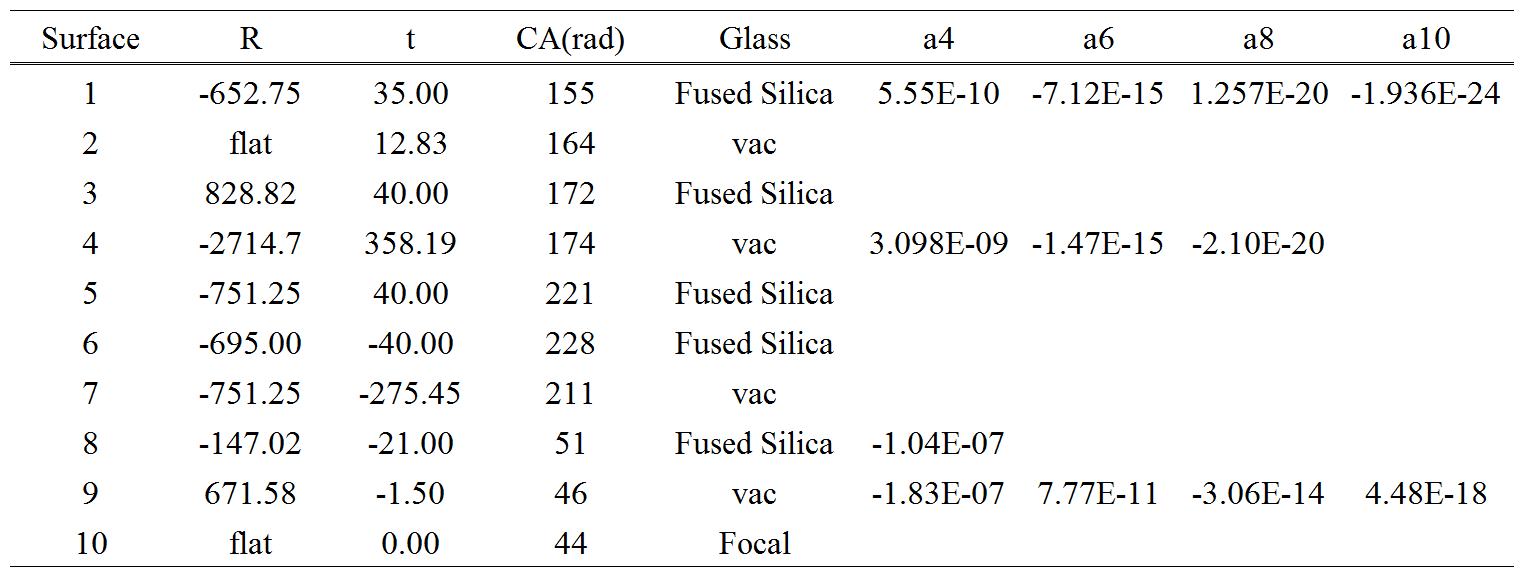}
\end{center}
\end{table*}

%-------------
%-------------

The necessity to use a very fast camera, dictated by the telescope aperture and available pixel sizes (15~$\mu$m in our case) very early drove us to a Schmidt optical design.  After some effort to produce a design with a single corrector element, a configuration with a doublet was chosen, both for performance reasons and manufacturability.  Each corrector has a strongly aspheric surface, but only one aspheric per element; a single corrector would have had to be thick and have strong aspheres on both surfaces. The corrector group has slight negative power, which offsets the chromatic errors in the positive field flattener. 

The active area of the detector is 61.4~mm square, and we use 60.6~mm of this to image the slit, which has a height of 139~mm, from 940~nm to 1.26~$\mu$m.  The final camera design has an effective focal length of almost exactly 300~mm, and the field angle at the corner of the image is 8.1 degrees. 

Early designs with a conventional first-surface mirror gave nearly acceptable performance, but the introduction of a mirror with a reflective surface on the rear improved the image quality markedly, even with spherical surfaces on both sides.  This is incorporated in the final design. The front surface of the mirror has slightly less curvature than the rear, so the mirror has slight positive refractive power.  A true Mangin mirror incorporates a negative meniscus lens to balance the undercorrected (negative) spherical aberration of the mirror surface.  In our case the negative corrector group produces a large amount of overcorrected (positive) spherical aberration, driving our Mangin mirror to have a positive meniscus lens, which adds additional negative spherical to the mirror surface itself to balance the overall system.

The field flattener lens has a very strongly aspheric rear surface which is slightly convex and whose vertex is only 1.5~mm from the detector.  The asphere only acts to remove zonal focus errors and has small effect on the best-focus images.  The final focal surface is acceptably flat (though slightly tilted (0.08~degrees in the dispersion direction), which is acceptable for a spectrograph) and has deviations from best focus of less than 10~$\mu$m everywhere except in the extreme corners of the field, where the focus errors rise to 20~$\mu$m at the extreme red and full image height. 

The camera design was optimized using Zemax, with RMS spot size as the merit function.  The camera was optimized as part of the whole spectrograph system, but the image aberrations are almost entirely due to the camera; the contributions from the Schmidt collimator are very small. A decision was taken early to optimize the collimator separately, so the beam delivered to the camera is parallel, and all the aberration correction is in the camera.  This is somewhat suboptimal, but greatly simplifies testing, especially since the collimator and spectrograph proper are being manufactured in France at Winlight Systems, and the NIR camera is being integrated in the U.S. at JHU.

\begin{figure}
   \begin{center}
   \begin{tabular}{c}
   \includegraphics[scale=0.45]{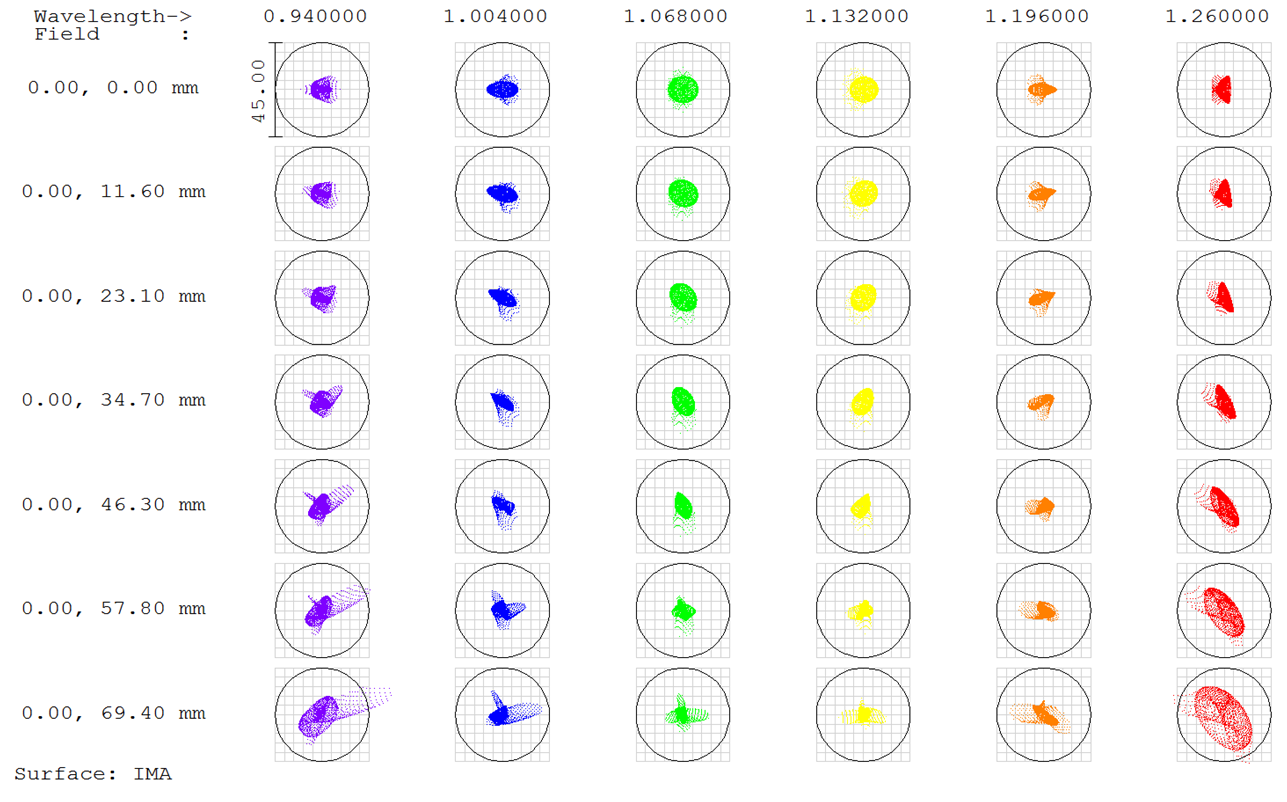}
   \end{tabular}
   \end{center}
   \caption[example] 
%>>>> use \label inside caption to get Fig. number with \ref{}
   { \label{fig:Fig4} 
Spot diagram for the near infrared camera.  The slit position, in millimeters, from the center of the slit at the top row, to the edge of the slit at the bottom row.  Columns represent different wavelength, in $\mu$m, and are labeled at the top of each column.  Each spot is referenced to a 45~$\mu$m diameter circle, which is 3 pixels.}
   \end{figure}

The design performance is quite good.  The RMS image diameters are at worst 22~$\mu$m in the corners, and are 15 or less almost everywhere in the field.  A spot diagram is shown in Figure \ref{fig:Fig4}.  The fiber footprint is 40.5~$\mu$m in diameter, so the aberrations add little to the geometric image size.  The image convolved with the fiber has RMS diameter which varies from 44~$\mu$m to 48~$\mu$m over the whole field.  The far wings of the image are very important for this instrument both because we wish to avoid crosstalk between adjacent fiber spectra (which are 98~$\mu$m apart on the detector) and because the infrared sky is dominated by very strong OH emission lines which will contaminate the faint object spectra.  The 98\% encircled energy diameter for this design varies from 76 to 88~$\mu$m and is less than 80 almost everywhere, and the width of the strip along the dispersion direction which contains 99\% of the energy is less than the fiber separation except, again, in the far red corner, where it rises to 99~$\mu$m.  This is a crude measure of crosstalk, and a more careful integral treatment gives crosstalk numbers of less than a tenth of a percent with so-called optimal extraction.  Again, performance like this is necessary because of the strong OH lines.

The NIR camera optics incorporate pair of coatings to greatly reduce the out-of-band light reaching the detector.  The rear reflecting surface of the Mangin mirror is a dichroic, which reflects in-band wavelengths and transmits wavelengths beyond 1.26~$\mu$m.   The rear surface of the second corrector element contains a thermal rejection coating.  This rather complex coating system is necessitated by the fact that we must use a detector with a 1.7~$\mu$m cutoff because stable HgCdTe material with shorter cutoffs cannot be produced, and the very low backgrounds necessary for faint-object spectroscopy demand that we efficiently reject the thermal radiation from the warm spectrograph, both entering in the beam and scattered. This is discussed more fully in the Thermal Background section of this paper.

%-------------
%-------------

%%%%%%%%%%%%%%%%%%%%%%%%%%%%%%%%%%%%%%%%%%%%%%%%%%%%%%%%%%%%%
\section{OPTOMECHANICAL DESIGN}
\label{sec:OPTOMECHANICAL_DESIGN} 

The optical prescription for this fast Schmidt design requires that the optics be located with precision, and this is challenging, for several reasons.  First, the optics will be mounted and aligned at room temperature but must operate at temperatures that are, in some cases, well below ambient.  Second, with the exception of the field flattener, the optics are large, and heavy.  The corrector elements are approximately 375~mm in diameter, with the first element having a mass of 11~kg, and the second element having a mass of 7~kg.  The primary mirror diameter is just under 505~mm with a mass of 17~kg.  And third, the vibration signature of the cyrocooler requires that the optical mounts be stiff to avoid resonances below 120~Hz.  In the design presented here, we have tried hard to account for these considerations, as well as the alignment tolerances discussed below.

\begin{figure}[H]
   \begin{center}
   \begin{tabular}{c}
   \includegraphics[scale=0.6]{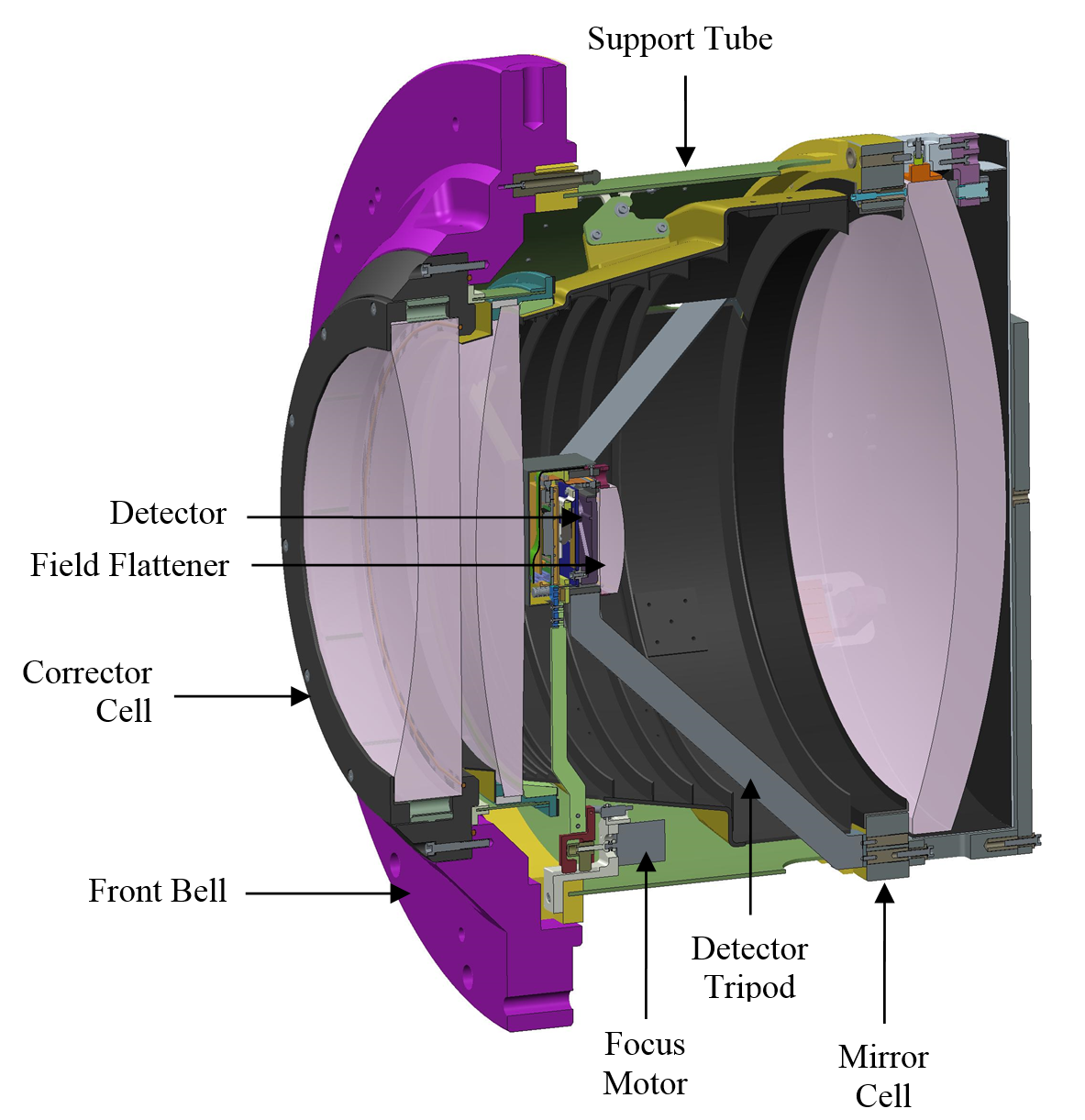}
   \end{tabular}
   \end{center}
   \caption[example] 
%>>>> use \label inside caption to get Fig. number with \ref{}
   { \label{fig:Fig5} 
NIR camera optomechanics.  The front bell is the principal optomechanical reference, locating the corrector cell and the primary mirror cell, which is supported at the end of a composite tube.  The field flattener is mounted to the detector mounting frame, which is metered with respect to the mirror cell by a silicon carbide detector tripod.}
   \end{figure}

The optomechanics of the camera, i.e. the components that hold and locate the optics, are shown in Figure \ref{fig:Fig5}.  Here, the front bell is the principal optomechanical reference, a reasonable choice given its size and mass.  The corrector is located, without adjustment, to a reference bore and shoulder machined into the front side of the bell.  The bore and shoulder {\it define} the optical axis.  The mirror and its mount are attached to the end of a composite support tube cantilevered off the rear side of the bell.  This provides the stiffness required to meet the vibration requirement and the thermal isolation needed to minimize cooler power.  The field flattener and detector are located by a silicon carbide detector tripod, which is attached to the face of the mirror cell.  Silicon carbide has a very low coefficient of thermal expansion (CTE), close to fused silica, and most importantly, has a very high thermal conductivity, which is needed for cooling the detector.  The radiation shield attached to the inside wall of the support tube serves as a light baffle.  This arrangement leads to what are really two subassemblies: the {\it front bell assembly}, containing the corrector; and the {\it primary optomechanical assembly}, which consists of the mirror cell, detector tripod, field flattener, detector, radiation shield, and focus mechanism.  Once assembled the two subassemblies are attached and aligned to achieve the required placement tolerances. 

What follows is a description of the various subassemblies that comprise the optomechanics for the camera.  A discussion of the optomechanical tolerances is presented first, since it is the placement tolerances that drives much of the design.  This section concludes with a discussion of the dynamic structural stability of the design.

%%-----------------------------------------------------------
\subsection{Optomechanical tolerances}
\label{sec:Optomechanical_tolerances}

Table~\ref{tab:Tab2} summarizes the alignment tolerances for the optical design, excluding the detector.  The tolerances were determined first from sensitivity analyses conducted using Zemax, then tweaked based on the results from Monte Carlo analysis.  Based on the results of the Monte Carlo runs, we do not expect the average RMS spot radius to increase more than 2~$\mu$m for the tolerances shown in the table.

%-------------
%-------------

Certainly we can rely on precision machining to achieve some of the tolerances.  The {\it tilts} for example are not daunting, with the exception of the mirror, and should be readily achieved by machining precision alone.

For the mirror, the tilt must be within 24~$\mu$m over the 490~mm diameter.  Grinding the face flat on the mirror could consume this entire tolerance, or more.  Thus tip and tilt adjustment of the primary is essential and will be set using optical feedback during alignment, if necessary.

As for the {\it decenters}, some are achievable without adjustment.  For example, placement of the first corrector element on the optical axis to within 150~$\mu$m is easily achievable, based on our experience with similar optics, i.e. the dual-element window in the FourStar~\cite{2013PASP..125..654P} infrared camera on Magellan.  The second element is more challenging.  It must be placed within 50~$\mu$m.  However we have devised a design, which is described below, that will allow us to achieve this tolerance through precision machining alone.  Hence, the corrector is placed without adjustment in this design.

\begin{table*}
\begin{center}
\caption{Optomechanical placement tolerances for the near infrared camera.\label{tab:Tab2}}
\includegraphics[scale=0.45]{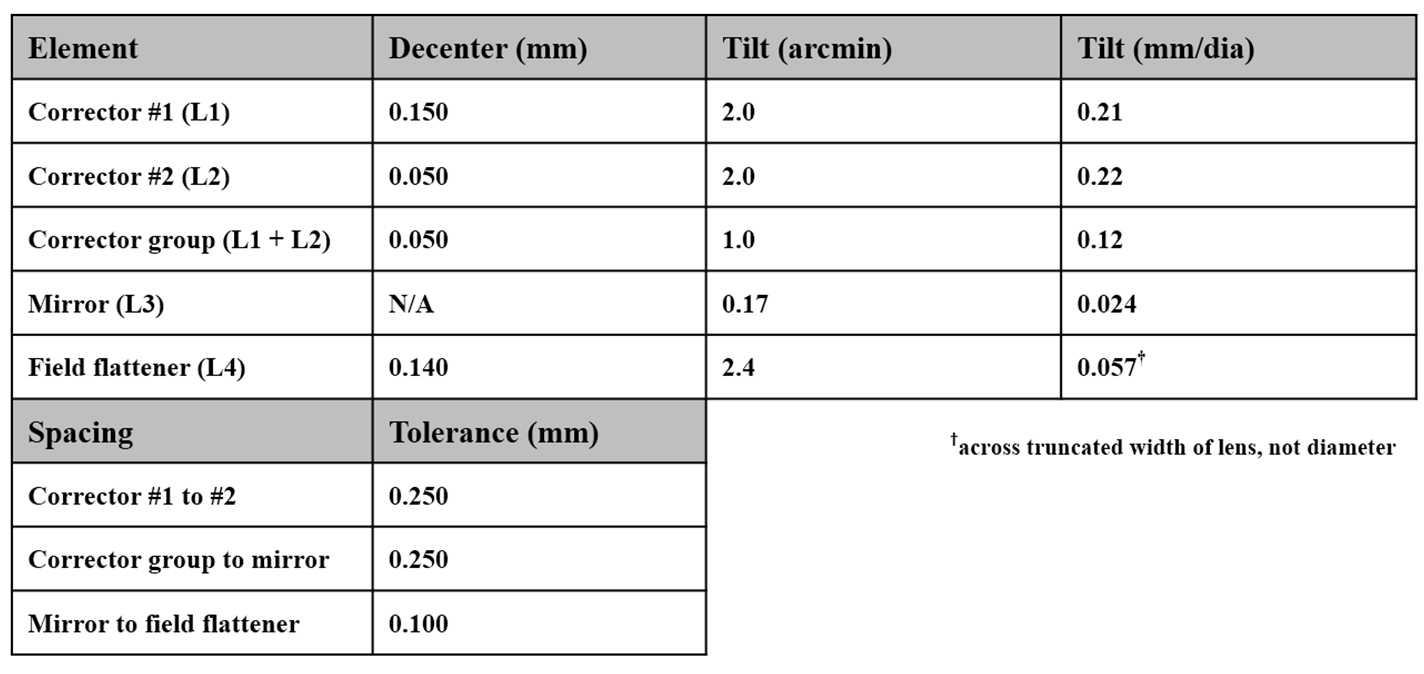}
\end{center}
\end{table*}

The mirror and field flattener are another matter.  Both are numerous interfaces away from the corrector.  It is difficult to imagine placing the mirror on axis to the accuracy required ($\sim$~90~$\mu$m) without adjustment.  Given the stiffness of the composite support tube, the gravity sag alone is approximately 250~$\mu$m.  Therefore we will adjust the position of the mirror to achieve the accuracy required, and the scheme for doing this is described below.

The decenter tolerance for the field flattener is 140~$\mu$m and is reasonable.  Given the numerous interfaces that ultimately define its location necessitates some form of adjustment to place the optic on-axis.

The tolerances on the placement of the optics along the optical axis, i.e. the {\it spacings}, are reasonable with the exception of the distance between the mirror and the field flattener, which must be held to 100~$\mu$m.  With the detector tripod being a bonded structure, plus the typical uncertainties in optical fabrication, some form of adjustment will be required to establish the relative distance between the mirror and the field flattener.  Here we have opted to piston the primary within its cell. The details are described below.  Once the mirror is adjusted relative to the field flattener, we will trim the length of the support tube to final length accounting for the as-built dimensions of the corrector, front bell, and mirror mount.

Accurate placement of the detector is critical to achieving the desired performance.  At f/1.07 the detector must be located to within 10~$\mu$m of the focal surface.  This cannot be reasonably done without adjustment in piston, tip and tilt.

To summarize, based on the factors discussed above, we have adopted the following adjustment philosophy:

1) {\it The corrector} - The corrector elements are to be mounted in a cell without provision for adjustment.  The cell is designed to ensure accurate centration, spacing, and tilt of the two elements relative to each other and to the reference surfaces that locate the cell within the front bell.  The bore in the front bell that accepts the corrector cell will establish the optical axis.

2) {\it The mirror} - The mirror will be adjustable in tip, tilt, piston, x and y.  Design features in the primary cell facilitate these five adjustments.  Placement of the mirror with respect to the corrector is achieved by trimming the length of the support tube, once the field flattener is spaced; hence the support tube serves as a shim.

3) {\it The field flattener} - The field flattener will be adjustable in x and y only by translating the support tube, i.e the entire primary optomechanical assembly.  To establish tip and tilt we will rely on a precision bonding fixture to locate the lens in its mount.  The spacing to the mirror will be adjusted by piston of the primary.

4) {\it The detector} – The detector will be adjusted using a focus mechanism that allows for piston, tip and tilt over a small range of motion.

%%-----------------------------------------------------------
\subsection{Corrector mount}
\label{sec:Corrector_mount} 

\begin{figure}
   \begin{center}
   \begin{tabular}{c}
   \includegraphics[scale=0.41]{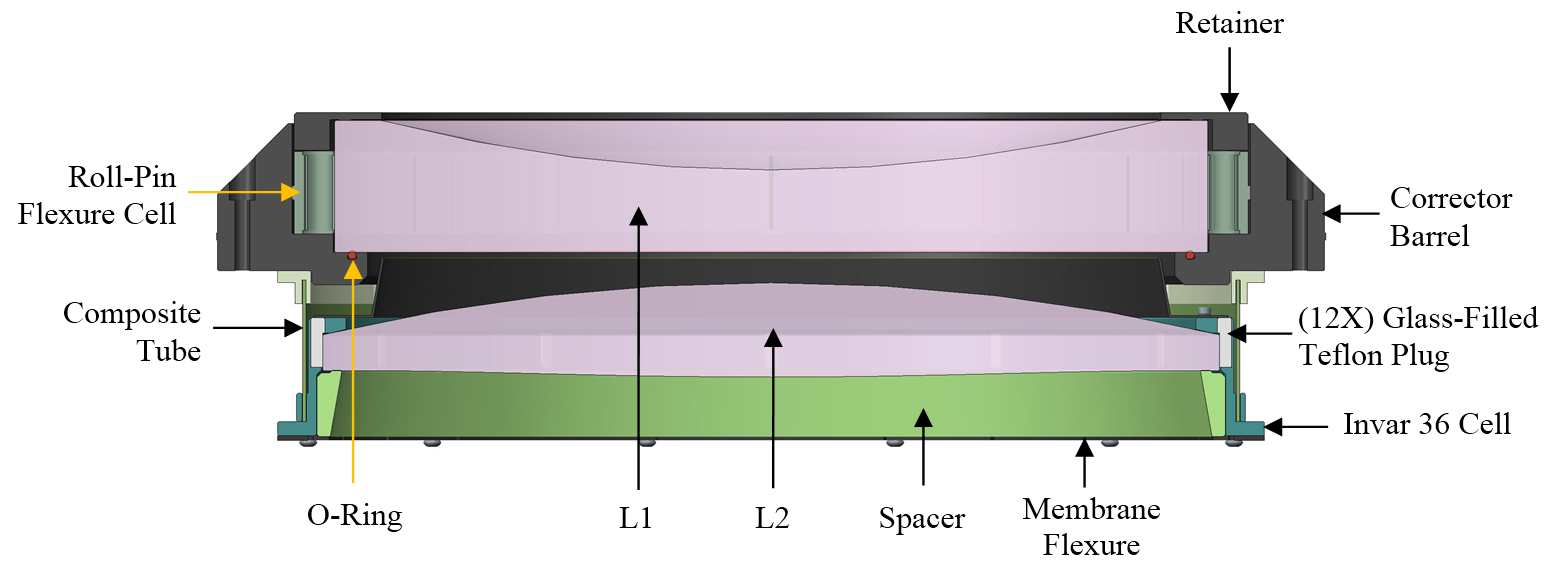}
   \end{tabular}
   \end{center}
   \caption[example] 
%>>>> use \label inside caption to get Fig. number with \ref{}
   { \label{fig:Fig6} 
Cross-section of the corrector assembly.  The dual-element corrector cell.  The first element, L1, serves as the vacuum window, and is centered by a roll-pin flexure cell inserted in the corrector barrel.  The compliant flexures accommodate the differential contraction between the fused silica lens and the aluminum cell. An O-ring at the base of the lens provides the vacuum seal.  The second corrector element, L2, also fused silica, is mounted in an athermal invar cell attached to the rear of the corrector barrel.  Retention rings with integral flexures constrain the lenses along the axis and allow for differential contraction in the axial direction.}
   \end{figure}

The dual-element corrector cell is shown in Figure \ref{fig:Fig6}.  The first element is centered by an aluminum roll-pin flexure cell~\cite{2010SPIE.7739E.119S} inserted into the front of the aluminum corrector barrel.  Compliant flexures within the cell accommodate differential radial expansion/contraction between the cell and the fused silica lens as the assembly is cooled from an ambient lab environment to the observatory operating temperature of 5~C.  A Viton O-ring at the base of the barrel seals against the flat on the rear of the lens.  A retaining ring compresses the O-ring, seating the lens against the shoulder at the base of the barrel.  The ring is fastened to the cell with shoulder bolts and belleville washer stacks to ensure proper preload and compliance to accommodate differential thermal expansion/contraction.  The use of aluminum to mount the first element is reasonable.  Aluminum is easy to machine, it matches the front bell to which it is attached, and the thermal excursion is small, about 30~C worst case; hence the required cell compliance is small enough to allow its use.

The second corrector element, also fused silica, is mounted in a hybrid athermal cell, an invar ring bonded to a composite tube.  A set of 12 glass filled Teflon plugs pressed into the ring athermalize the cell, the plugs being sized such that the effective CTE of the cell matches the fused silica lens.  The plugs are bored to fit the lens after being pressed into the invar ring.  The 1.5~mm thick composite tube thermally isolates the lens from the corrector barrel.  The use of invar aides in this regard as well given its low thermal conductivity.  The total mass of the cell assembly is 33.5~kg.

%-------------
%-------------

%%-----------------------------------------------------------
\subsection{Mangin mirror mount}
\label{sec:Mangin_mirror_mount} 

\begin{figure}
   \begin{center}
   \begin{tabular}{c}
   \includegraphics[scale=0.5]{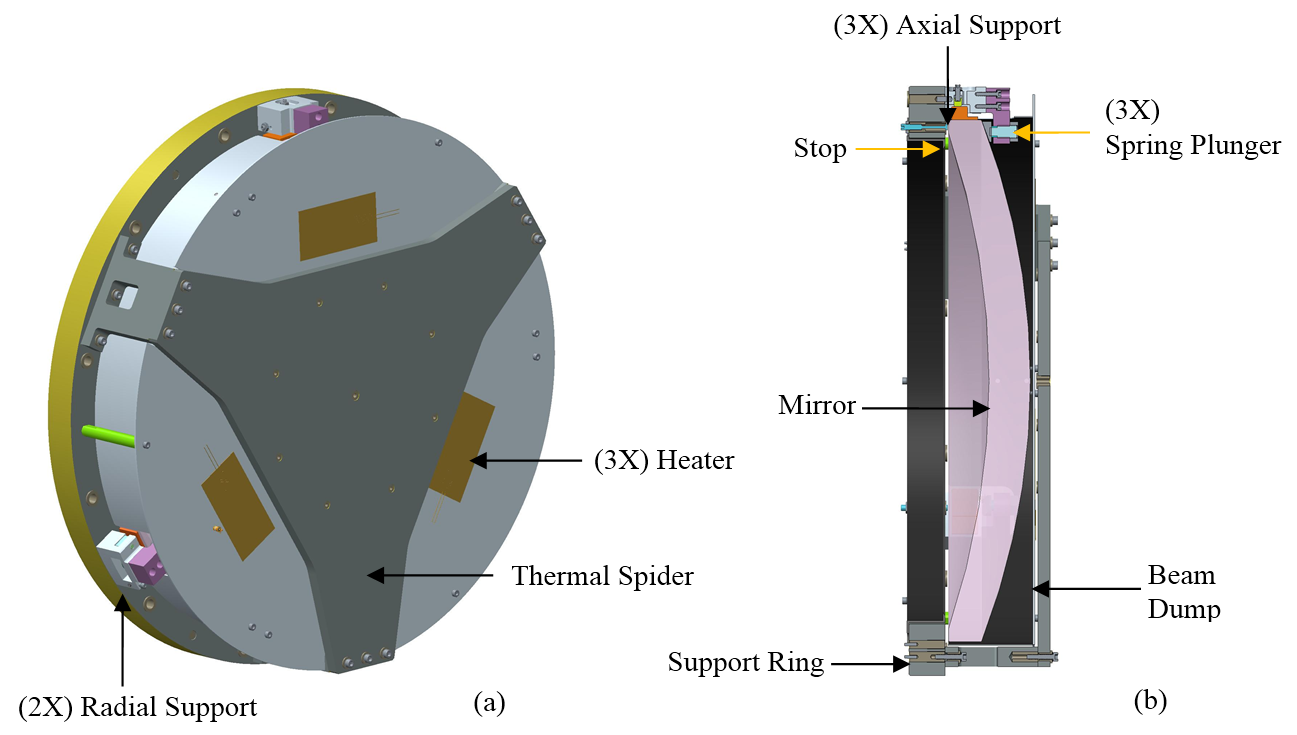}
   \end{tabular}
   \end{center}
   \caption[example] 
%>>>> use \label inside caption to get Fig. number with \ref{}
   { \label{fig:Fig7} 
Mangin mirror mount:  (a) isometric view and (b) cross section.  The mirror is supported off the back of a silicon carbide support ring.  It rests on two adjustable radial supports, each support being 30 degrees off the horizontal axis.  Three adjustable post flexures support the mirror on-axis and provide tip, tilt, and piston adjustment.  A beam dump behind the mirror thermally floats on G10 standoffs and collects light at wavelengths beyond the mirror coating cutoff around 1.28~$\mu$m.}
   \end{figure}

The Mangin mirror is mounted to the back of a silicon carbide support ring; see Figure \ref{fig:Fig7}.  The mirror rests on two composite-tired bearings housed in support blocks bolted to the back of the ring.  The use of bearings thermally isolates the mirror from the silicon carbide ring, which runs very cold since it is part of the thermal path to cool the detector.  The G10 tire pressed over the outer race softens the contact with the mirror and improves thermal isolation.  Each support is 30 degrees off the horizontal.  A third, upper support block, constrains the mirror in rotation.  Invar pads bonded to the circumference of the mirror at three support locations eliminate glass to metal line/point contact.  

The two lower supports can be adjusted in the radial direction to center the mirror on axis.  A set screw in each block adjusts the radial position of the bearing.  Jacking both screws inward raises the mirror in the vertical direction.  Outward lowers the mirror.  Jacking the left screw in and the right screw out by an equal amount shifts the mirror to the right. The opposite shifts it to the left.

%-------------
%-------------

Placement of the mirror along the optical axis is established by three titanium post flexures.  The flexures thread into the support ring and protrude slightly from the rear surface.  The flat on the face of the mirror rests on these three points.  Thin invar pads between the tip of the flexure and the mirror eliminate the metal-to-glass point contact.  Spring plungers, one behind the mirror at each location, push against invar pads bonded to the back the mirror and register the mirror against flexures.  The flexures are adjustable from the front side to enable piston, tip, and tilt of the mirror.

An optical stop fastened to the front of the ring clips stray light outside the clear aperture.  An aluminum beam dump behind the mirror absorbs radiation longward of 1.26~$\mu$m, which passes through the dichroic mirror coating.  The inner surface is lined with Acktar Metal Velvet foil and its rear surface is polished for low emissivity.  Aside from collecting unwanted radiation, the beam dump also provides a constant-temperature surface to which the mirror radiates.  This mitigates the possibility of gradients in the mirror from view factors to multiple surfaces having different temperature.  Three 5~W heaters on the rear of the beam dump aid in warming the camera to ambient from a cold state.

Behind the beam dump, and attached to the mirror support ring, is the thermal spider, which thermally connects the mirror support ring to the detector cyrocooler.  The spider is supported off the back of the ring by three standoffs.  The support ring, tripod legs, thermal spider, and standoffs are all made from SC-1000 silicon carbide from Kyocera\footnote{Kyocera Industrial Ceramics Corp., www.kyocera.com}.

The entire assembly has a mass of 36~kg.

%%-----------------------------------------------------------
\subsection{Field flattener mount}
\label{sec:Field_flattener_mount} 

Mounting the field flattener is a challenge.  There is very little material outside the clear aperture to hold the lens and there is no room for adjustments.  In addition, the only good reference for placing the optic on axis is the four segments of the outer diameter that remain after grinding the flats.  We have considered bonding the lens directly to the detector mounting frame but have backed away from the idea.  It is risky, and for the alignment procedure we envision, the lens needs to be removable.

\begin{figure}
   \begin{center}
   \begin{tabular}{c}
   \includegraphics[scale=0.35]{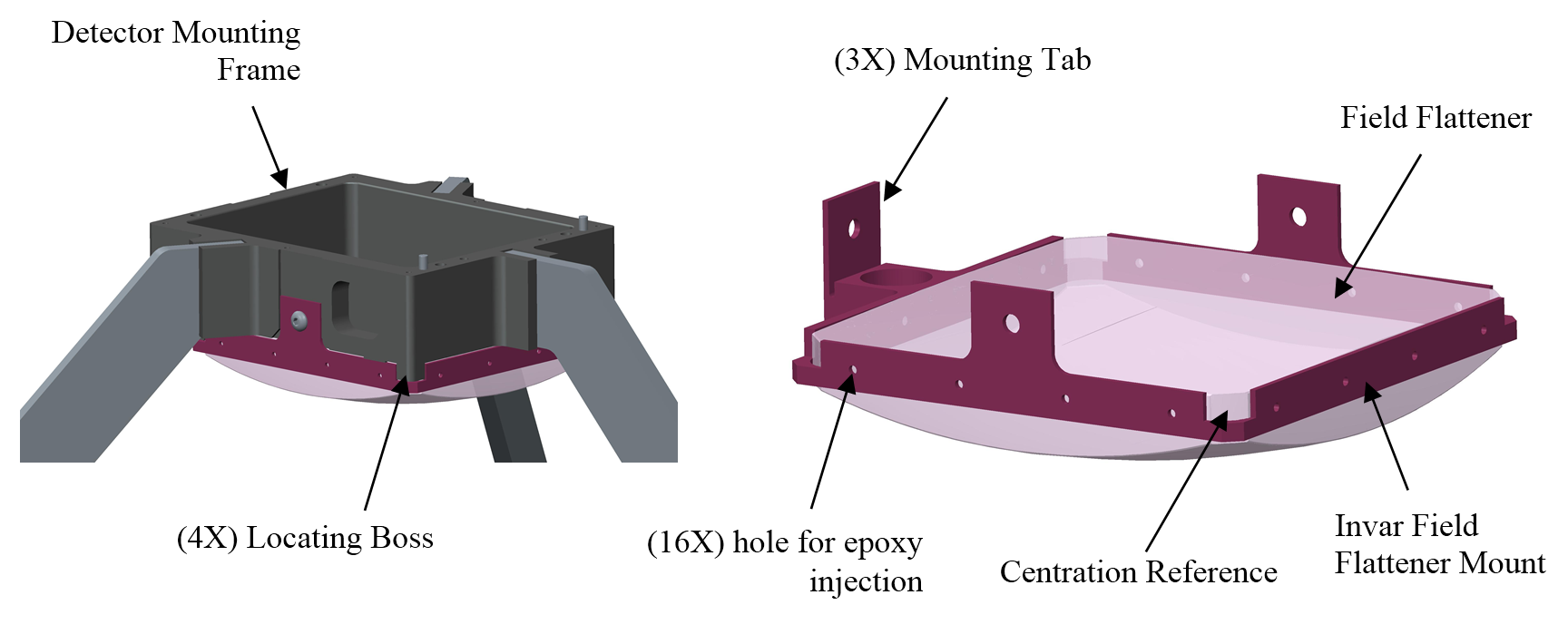}
   \end{tabular}
   \end{center}
   \caption[example] 
%>>>> use \label inside caption to get Fig. number with \ref{}
   { \label{fig:Fig8} 
Field flattener mount details.  The field flattener is precision bonded to an Invar 36 mount with three tabs that fix the assembly to the detector mounting frame.  The lens is centered using the four exposed corners, which engage four tabs protruding from the bottom of the detector mounting frame.  The lip of the mount seats against three bosses on the bottom of the detector mounting frame, locating the optic axially.}
   \end{figure}

The design we have settled on is shown in Figure \ref{fig:Fig8}.  The field flattener is precision bonded to an invar mount with three integral tabs to attach it to the detector frame.  The four corners of the mount are relieved leaving the round segments of the lens exposed.  These round segments engage four locating bosses protruding from the bottom of the mounting frame, thus centering the lens.  The bosses will be machined to fit the lens diameter with appropriate clearance.  Three separate, and much shorter, protrusions extending from the bottom of the detector mounting frame will register the lip of the field flattener mount, locating the lens axially.

%-------------
%-------------

Clearly, the mount, the detector mounting frame, and the bond fixture must be very precise.  However we believe this is possible through the use of wire electrical discharge machining (wire EDM).  We will machine the perimeter of both the mount and detector mounting frame using wire EDM and we should be able to achieve dimensional precision of order 20 $\mu$m or better.  If needed, we will match each field flattener mount to the lens.

\begin{figure}
   \begin{center}
   \begin{tabular}{c}
   \includegraphics[scale=0.5]{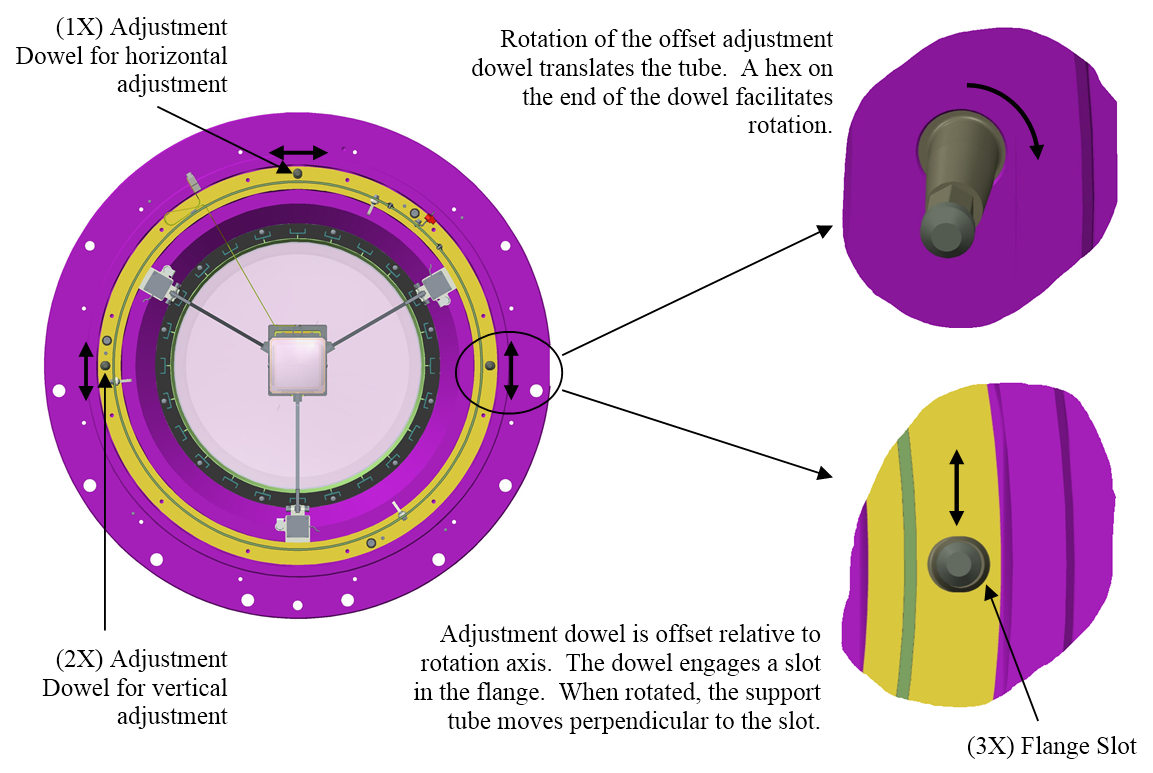}
   \end{tabular}
   \end{center}
   \caption[example] 
%>>>> use \label inside caption to get Fig. number with \ref{}
   { \label{fig:Fig9} 
Field flattener alignment details.  The image on the left shows the support tube interface to the front bell; as seen from the mirror.  Three slots in the tube flange engage three adjustment dowels.  The dowels have a simple circular cam profile such that rotation of the dowel translates the tube in a direction perpendicular to the axis of the slot.  Rotation of the two horizontal dowels shifts the tube vertically.  Rotation of the upper dowel shifts the tube horizontally.  Locking screws (not visible) lock the dowel to the bell once adjusted.}
   \end{figure}

To adjust the field flattener we will shift the entire primary optical assembly by translating the support tube at the interface between the tube and front bell.  The mechanism for doing this is shown in Figure \ref{fig:Fig9}.  Three adjustment dowels located in the front bell engage three radially aligned slots in the front flange of the tube.  The adjustment dowels have two diameters: one that engages the front bell, and a smaller diameter, off-axis relative to the first, that engages the slot in the flange.  Two of the slots are aligned with the horizontal axis and one with the vertical.  With the flange bolts loosened, rotation of the two dowels on the horizontal axis shifts the tube in the vertical direction, without affecting the horizontal location.  Similarly, rotation of the dowel on the vertical axis shifts the tube in the horizontal direction.  A locking screw buried in the center of each dowel locks the adjustment.  Once the field flattener is positioned on axis, three floating alignment pins, not shown in the figure, are tightened, and provide a permanent reference for accurate/repeatable placement of the support tube should it need to be removed.

%-------------
%-------------

%%-----------------------------------------------------------
\subsection{Detector mount}
\label{sec:Detector_mount} 

\begin{figure}
   \begin{center}
   \begin{tabular}{c}
   \includegraphics[scale=0.5]{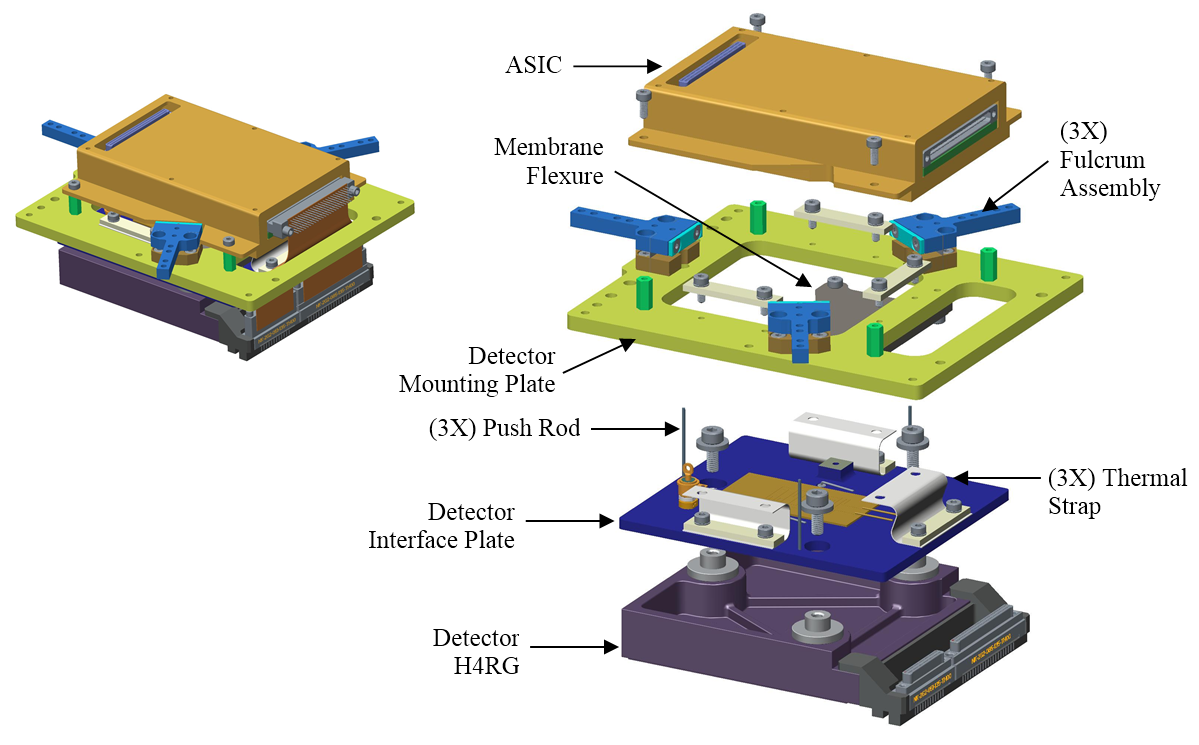}
   \end{tabular}
   \end{center}
   \caption[example] 
%>>>> use \label inside caption to get Fig. number with \ref{}
   { \label{fig:Fig10} 
Detector assembly design details.  The detector is kinematically located and attached to the detector interface plate, which is suspended from the end of a bronze membrane flexure.  This flexure constrains the detector in rotation but leaves it free to execute small motions in piston, tip, and tilt.  Three push rods, which are part of the focus mechanism extend perpendicular to the detector interface plate and are clamped to the three fulcrum assemblies, also part of the focus mechanism.  The ASIC mounts on standoffs above the detector mounting plate.}
   \end{figure}

The detector assembly, shown in Figure \ref{fig:Fig10}, is mounted to the top surface of the detector mounting frame, which is bonded to the top of the detector tripod.  This surface is the reference surface for the detector; the focal plane is a known distance in front of this surface.  The detector assembly interface to the frame is the detector mounting plate, which is also made of invar.  The plate is located on the frame by two alignment dowels and is fastened by six screws.  During installation, two guide posts (not shown in the figure) are threaded into the mounting plate for the purpose of guiding the assembly safely into the box; once assembled the posts are removed.

%-------------
%-------------

Details of the detector assembly are shown in the exploded view in Figure \ref{fig:Fig10}.  The detector is kinematically located and attached to the detector interface plate, which is suspended from the end of a bronze membrane flexure.  This flexure constrains the detector in rotation but leaves it free to execute small motions in piston, tip, and tilt.  Three push rods, which are part of the focus mechanism, extend perpendicular to the detector interface plate and are clamped to the three fulcrum assemblies, also part of the focus mechanism.  Motion of the push rods affects piston, tip, and tilt adjustments of the detector.  The focus mechanism is described in more detail below.  The cold readout electronics, the Teledyne ASIC (Application-Specific Integrated Circuit), mounts on standoffs above the detector mounting plate and a short flex cable connects the ASIC to the detector. 
 
The detector is cooled largely by radiation to the cold field flattener in front of the detector, and to a lesser degree, by conduction through three silver straps that couple the detector interface plate to the detector mounting plate, which is intimately connected thermally to the detector mounting frame.  Detector temperature is controlled by a 2~W resistive Kapton film heater bonded to the back of the detector interface plate.  Temperature feedback is provided by two platinum RTDs bonded to the plate; only one is needed however an additional sensor is added for redundancy.  A Klixon 3BT-6 thermostat is also attached to the interface plate.  The normally closed thermostat is wired in series with the heater and opens if the temperature exceeds 40~C.

Some effort has been invested in making this design as compact as possible to minimize the amount of light obscured by the hardware.  The Teledyne H4RG-15 is not square but has a protrusion on one side to handle the output connectors.  This is unfortunate in that it costs light.  Given this geometry and the geometry of the detector mount, we can calculate the obscuration this represents in the input beam.  Clearly, as one goes off-axis, the obscuration increases, because one begins to look at both the detector box and the tripod legs obliquely instead of face-on.  The camera has a full field of about 16~degrees, so this effect is not small.  On-axis the obscuring area is 15\%.  This rises to 17\% at full field.

%%-----------------------------------------------------------
\subsection{Focus mechanism}
\label{sec:Focus_mechanism} 

At f/1.07 the depth of focus for this camera is very small, +/-~10~$\mu$m, and it is very unlikely we could achieve acceptable focus, in a reasonable amount of time, by simply shimming the detector.  Therefore focus will be set using an active focus mechanism.  As described above, the detector is suspended from a bronze membrane flexure that constrains rotation but allows small piston, tip and tilt motions.  These motions are affected by translating three push rods attached to the detector interface plate.  Each push rod is coupled to one end of a fulcrum assembly; the fulcrum being a thin flexure.  On the other side of the fulcrum a long narrow lever is attached.  With three levers like this, the piston, tip and tilt of the detector can be adjusted for focus, evaluating optical performance, taking up any changes caused by changing environmental temperature, disassembly/reassembly, or any other less than perfectly understood environmental/setup conditions.

\begin{figure}
   \begin{center}
   \begin{tabular}{c}
   \includegraphics[scale=0.45]{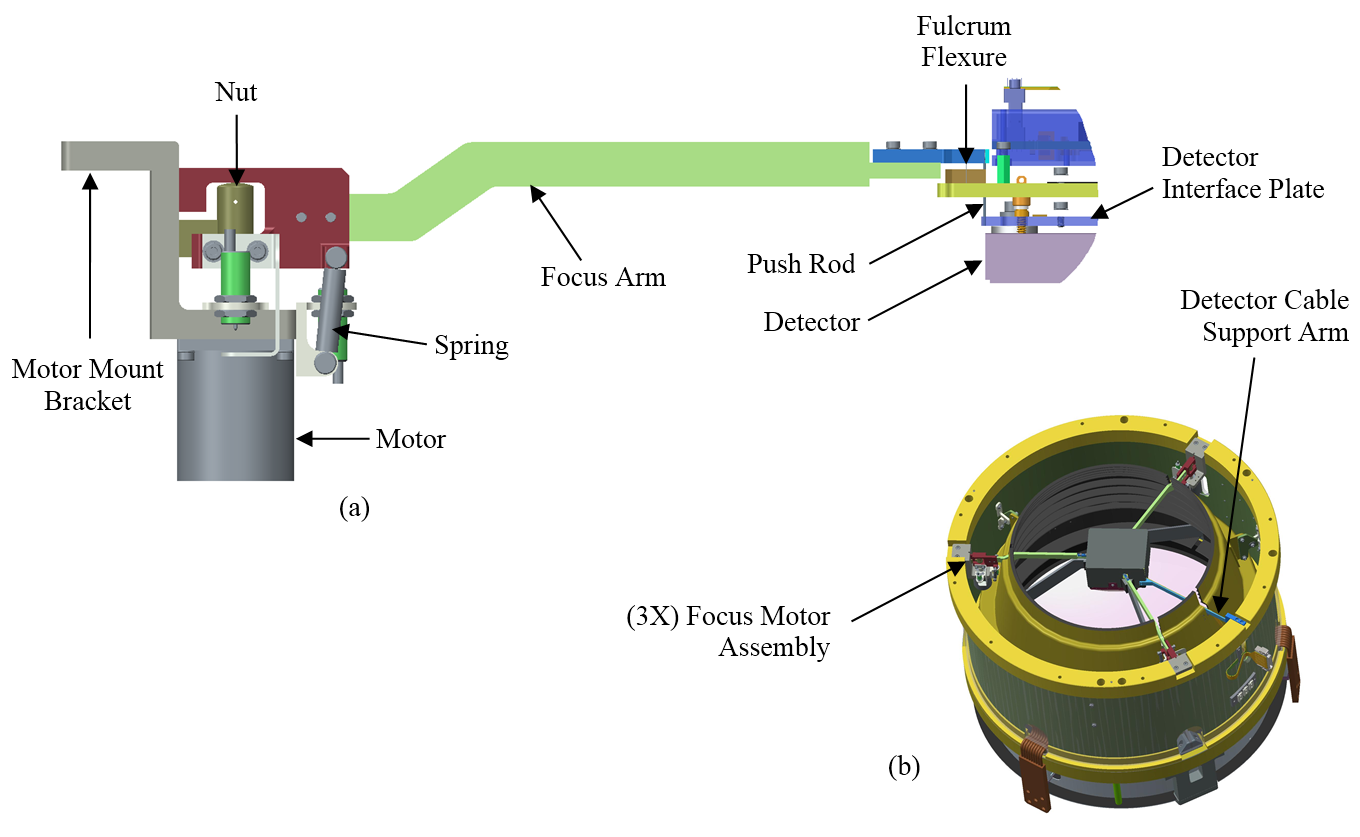}
   \end{tabular}
   \end{center}
   \caption[example] 
%>>>> use \label inside caption to get Fig. number with \ref{}
   { \label{fig:Fig11} 
(a) Focus mechanism design details, and (b) view of the three focus motor assemblies mounted to the front ring of the support tube.  Rotation of the focus motor translates the nut causing the focus arm to pivot about the focus fulcrum, which in turn translates the push rod attached to the detector interface plate.  The mechanism provides a large mechanical reduction, about 38:1, defined by the distance from the nut to the focus fulcrum divided by the distance from the focus fulcrum to the push rod.}
   \end{figure}

Design details for the mechanism are shown in Figure \ref{fig:Fig11}.  The focus arm is translated by a motorized, fine-pitched, lead-screw and nut.  The mechanism provides a large mechanical reduction, about 38:1, defined by the distance from the nut to the focus fulcrum divided by the distance from the focus fulcrum to the push rod.  For a 200 step motor and a 0.5~mm pitch lead-screw, the expected resolution is 65~nm.  The mechanism has been designed for a total range of motion of +/-~150~$\mu$m.  As shown in the figure, the three focus arms are aligned with the three tripod struts.  This reduces beam obscuration.

%-------------
%-------------

The focus motor assembly is a simple design.  A Phytron VSS 32 vacuum-compatible motor is supported by a bracket that mounts to the front of the support tube flange.  A 17-4 PH stainless steel lead-screw (0.5~mm pitch) directly attached to the motor drives a Nitronic 60 nut (both of which will be coated with tungsten disulfide).  The nut is constrained in rotation by a tab that engages a slot in the mounting bracket.  Hence when the motor turns, the nut translates.  Two extension springs preload the focus arm against the tip of the nut, and the arm translates in unison with the nut.  Two limit switches limit the range of travel.  One of the switches also serves as a home sensor.

%%-----------------------------------------------------------
\subsection{Dynamic performance}
\label{Dynamic_performance} 

A significant consideration in the design of this camera is the impact that cryocooler vibration has on the optomechanical stability of the camera.  Considerable effort has gone into understanding the issue, i.e., the source of the problem, as well as how to mitigate it, and analysis to predict the impact; far too much to report here.  Therefore we limit the discussion here to a summary of the work done to mitigate the problem and discuss the numerical effort to predict the impact on performance.  For the interested reader, more detail on the Sunpower cyrocooler vibration, and the active damping control system developed to reduce it, can be found in a paper by Hope et al.\cite{Hope2014} (this conference).

As mentioned above, at the outset of the camera design effort, we set a requirement that the natural frequencies of structural elements in the load path extending from the coolers to the optics have natural frequencies in excess of 120~Hz, and if possible, in excess of 180~Hz.  This requirement also extended to the remainder of the spectrograph although we later relaxed the limit to 80~Hz for subsystems with no influence on optomechanical stability, or for subsystems where it could be demonstrated by analysis that a reduced limit was acceptable.  Finite element modeling was performed to determine mode frequencies and guide the design as it progressed, and later, once the design was sufficiently stiff, a frequency response analysis was conducted to determine the impact on the stability of the optics.  We summarize the results here.

The modal analysis was done in a tiered fashion given the complexity of the camera design.  A full finite element model of the camera, including the mount, was constructed with two simplifications: first, the mass of the detector and its mount were folded into the density of the detector mounting frame, and second, the mass of the components attached to the rear of the camera was folded into the density of the rear cover.  Separate detailed models were constructed for the detector mount and the rear cover assembly (with, coolers, pumps, etc.) to determine the modes of these subsystems.

Results for the full camera model showed the first mode of the system to be at 79~Hz, a rotational mode of the radiation shield.  The shield actually has two additional modes below 120~Hz.

To assess the optomechanical stability of the camera, i.e., the motion of the optics, a frequency response analysis (also called a sine sweep) was conducted using the same finite element model used for the camera modal analysis.  Of particular interest is the motion of the detector and the mirror.  Both were monitored in the analysis.  The model outputs displacements as a function of frequency for a harmonic forcing function.  Here the key parameter is the force amplitude applied by the coolers at the rear of the cryostat.

Considerable effort has been invested in trying to understand the force applied to the cryostat by the cooler; again, see Hope et al.\cite{Hope2014} (this conference).  The cooler itself sees forces from the internal moving pistons, and that net force is easy enough to determine by measuring the acceleration and knowing the mass; $force = mass \times acceleration$.  However, the forces are a function of frequency, operating power, and direction.  In addition, as a means of vibration mitigation the cooler itself is mounted in a compliant fashion with springs supporting the cooler on-axis and soft sorbothane off-axis, a detail we will likely change.  This, combined with a significant reduction in vibration from the active damping system, makes reliable measurement more difficult; in part because we are trying to measure accelerations near the noise floor of the accelerometer.

Nonetheless, based on accelerometer measurements taken with the cooler in its custom-designed compliant mount and mounted to a small test chamber, the forces acting on the cryostat with the cooler running at full power (240~W), and with the active damping on, were determined; see Table~\ref{tab:Tab3}.  

\begin{table*}
\begin{center}
\caption{Measured forces applied to the cryostat by the Sunpower GT running at 240~W with active damping.\label{tab:Tab3}}
\includegraphics[scale=0.3]{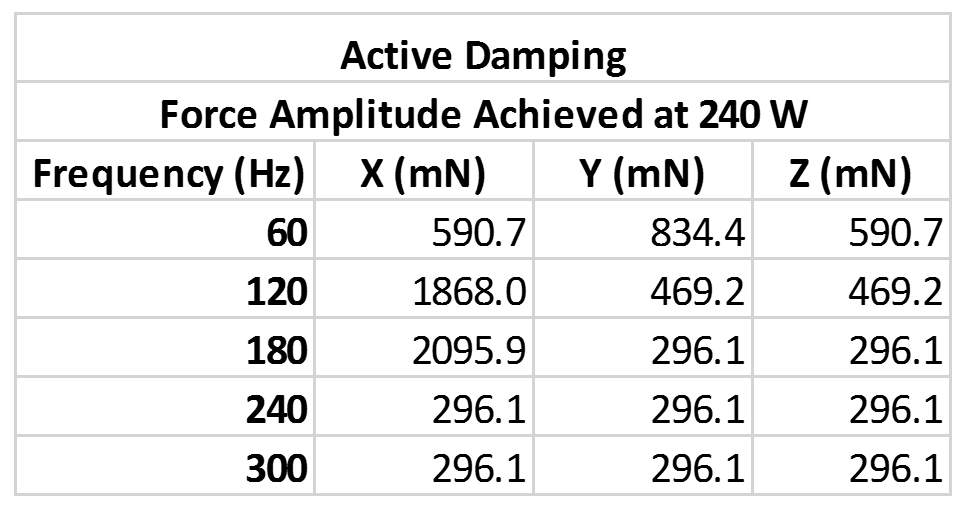}
\end{center}
\end{table*}

Using this data for the force amplitude in the frequency response analysis, the motions of the field flattener and mirror were determined.  The results indicate a displacement amplitude of less than 0.2~$\mu$m for the field flattener and less than 0.3~$\mu$m for the mirror.  In the case where the active damping is turned off, and the active damper is acting in a passive mode, the forces (not shown here) increase, and the displacement of the field flattener and mirror are $\sim$~0.7~$\mu$m and $\sim$ 1~$\mu$m, respectively.  Motions at these levels are small and will have no impact on the science.  And while it is possible that active damping is not needed we will implement it, just in case the predictions are wrong.

In an attempt to validate this modeling, we mounted a cooler in its compliant mount to a large test chamber, measured the acceleration, and compared the results with a frequency response model of the chamber.  The numerical and experimental results agreed within a factor of 2.  Considering the sensitivity of the model to boundary conditions, and the unfortunate fact that the test chamber just happens to have a resonance near 60~Hz, the agreement seems reasonable.

%%-----------------------------------------------------------
\subsection{Optomechanical alignment}
\label{sec:Optomechanical_alignment} 

The procedure envisioned for aligning the camera optics is as follows:

1) The front bell assembly and primary optical assembly will be assembled and mounted to a test stand on an optical bench such that the optical axis is horizontal.  In this configuration the corrector and field flattener will be removed.

2) In place of the field flattener will be a transparent surrogate alignment target having the same mount interface as the actual field flattener and a cross-hair on the nominal axis of the actual lens.  Similarly a transparent alignment plate will be installed in place of the corrector cell.  This target establishes the optical axis of the camera.

3) A theodolite will be placed on a two-axis stage (vertical and horizontal), on the optical bench upstream of the camera.  In autocollimation the theodolite will be set normal to the corrector target and centered on the target using the stage.  The theodolite axis is now the alignment reference for the camera optical axis.

4) With the bolts attaching the support tube to the front bell loosened slightly, the primary optical assembly will be adjusted laterally until the field flattener target is centered with the axis of the theodolite.  Given the cantilevered mass of the primary mirror we envision mounting a fixture to the rear of the cell with either spring-loaded or pneumatic feet to offload a large portion of the mass during this operation.

5) Lastly, the theodolite will be focused to the center of curvature of the rear (reflecting) surface of the primary, and the mirror will be adjusted, in-plane, until the center of curvature is on the theodolite axis.  Tip, tilt, and piston of the primary will be adjusted prior to installing the cell/tripod assembly to the composite tube, based on as-built metrology data from in-house Coordinate Measuring Machine (CMM) surveys of the mirror and cell/tripod assembly.

We will certainly want to verify the alignment at operating temperature.  To do that we will assemble the remaining cryostat components, cool the camera, and verify alignment of each target location.

%%%%%%%%%%%%%%%%%%%%%%%%%%%%%%%%%%%%%%%%%%%%%%%%%%%%%%%%%%%%%
\section{THERMAL DESIGN}
\label{sec:THERMAL_DESIGN} 

%%-----------------------------------------------------------
\subsection{Design details and equilibrium temperatures}
\label{sec:Design_details_and_equilibrium_temperatures} 

The camera thermal system is similar to most infrared cameras where the cameras are the cryostats.  One's first impression is that going to only 1.26~$\mu$m should not be sufficiently different from optical instruments that much care is necessary with thermal background, but that turns out not to be the case at all.  In fact, the construction of the camera in the dewar is actually key to achieving the desired performance.  It is to be noted that for this design, the whole camera assembly (mirror, detector, detector supports) is cold.  

For cooling, two Sunpower GT40 Stirling cycle coolers are used.  (Note, Subaru does not allow the use of LN2).  These devices have a {\it lift} (i.e. cooler capacity) of 15~W at 77~K, and consume approximately 260~W of power.  The free-piston design needs no external compressor.  Waste heat will be removed by a two-stage liquid cooling system and eventually dumped into the Subaru facility glycol system.  The unit is compact, and has a mean time to failure in excess of 100,000 hours ($\sim$~10 yrs).  The only real drawback, as discussed above, is vibration, which is mitigated by an active damping system.

There are two coupled thermal domains in this design.  The first cools the detector.  Heat is removed from the detector by the central cooler at the rear of the cryostat; see Figure \ref{fig:Fig2}.  Energy from the detector is conducted through the silicon carbide support tripod, to the silicon carbide mirror cell, and ultimately to the cooler through a set of copper wire-rope thermal straps that couple the cell to the cooler.  The second cools the radiation shield.  Energy incident on the shield -- largely form the corrector -- conducts through three copper wire-rope straps to a thermal pan suspended from the cryostat rear cover, to the second cooler, which is mounted off-axis.

In this design, the composite tube provides conductive isolation between the cold mirror and the ambient temperature front bell.  It also isolates the radiation shield, which is suspended from the internal diameter of the tube.  In a like manner, the G10 struts are used to isolate the thermal pan from the rear cover.

Thermal radiation between the cryostat walls and the cold internals is mitigated in the typical fashion. Multi-Layer Insulation (MLI) blankets are applied to cold surfaces with a direct view to the cryostat walls.  In addition the internal walls of the cryostat will be polished to a mirror finish.  For the inner surface of the rear cover, which is heavily lightweighted, a polished aluminum baffle plate covers the surface.  An MLI blanket is also present on the outer diameter of the radiation shield, to reduce loading from the inner diameter of the composite tube.  MLI blankets will be constructed with a total of 10 to 15 layers with 2~mil double-sided Vapor Deposited Aluminum (VDA) Mylar for the inner/outer layers, and 0.25~mil double-sided VDA Mylar for the internal layers; layers are separated by Dacron mesh.

\begin{figure}[H]
   \begin{center}
   \begin{tabular}{c}
   \includegraphics[scale=0.6]{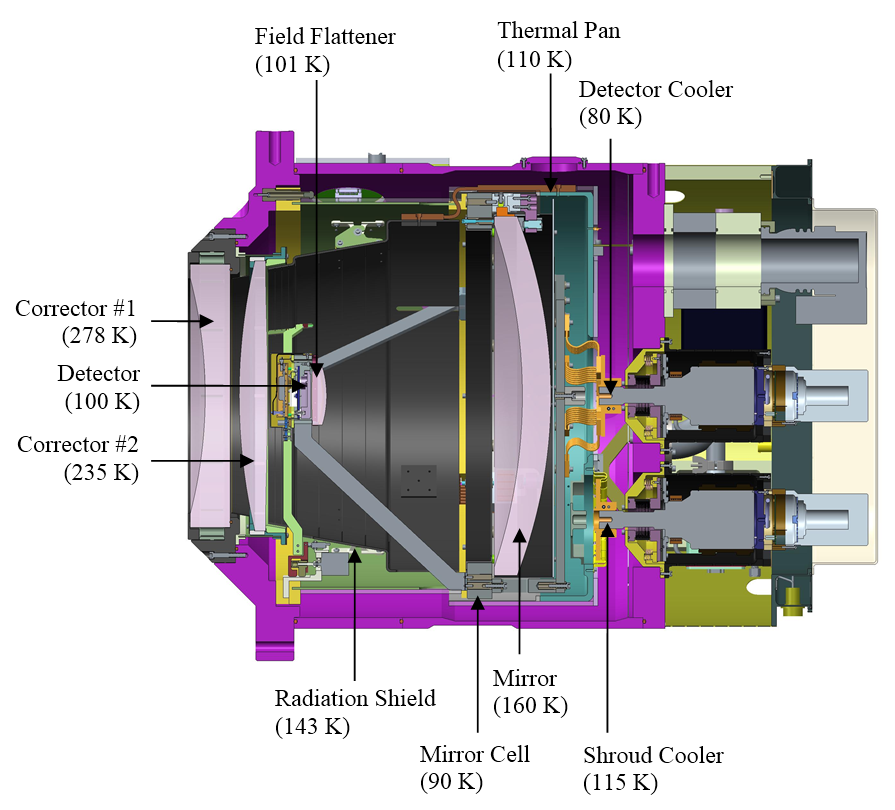}
   \end{tabular}
   \end{center}
   \caption[example] 
%>>>> use \label inside caption to get Fig. number with \ref{}
   { \label{fig:Fig12} 
Cross-sectional view of NIR camera showing thermal equilibrium temperatures.  The lowest foreseeable operating temperature for the detector, 100~K, requires that the mirror cell run at $\sim$ 90~K, and the detector cooler at $\sim$ 80~K.}
   \end{figure}

Steady state operating temperatures for the camera were determined using finite difference code and verified using finite element analysis.  Figure \ref{fig:Fig12} shows a cross section of the camera with component temperatures of interest.  For a detector temperature of 100~K, the lowest foreseeable operating temperature, the mirror cell must run at $\sim$ 90~K, and the detector cooler at $\sim$~80~K.  The radiation shield will be cooled to just over 140~K, low enough to minimize radiative loading on the detector tripod legs, and in turn, additional heat load on the detector cooler.  The thermal resistance between the shield and the shield cooler sets the operating temperature of the shield cooler at $\sim$ 115~K.  Radiative coupling to the corrector cools the rear inward facing element to $\sim$~240~K.  As discussed above, the support for the rear corrector element is designed to allow this element to float cold.  In so doing, the radial gradient in the first corrector lens, L1, is approximately 3~K center-to-edge, and for the second corrector lens, L2, the gradient is approximately 4~K; both are small enough to not adversely affect the transmitted optical wavefront.  The mirror floats at a temperature of 160~K with negligible gradeint center-to-edge.

%%-----------------------------------------------------------
\subsection{Heat loads}
\label{sec:Heat_loads}

The main heat load is radiation from the corrector (unlike the signal background, the thermal load is mostly at wavelengths of 10~$\mu$m or larger, at which the corrector is opaque and almost black), most of which $\sim$ 90~\%) strikes the radiation shield.  This is about 12~W.  Another $\sim$5~W will be lost through the MLI around the thermal shield and the back of the mirror, without which the thermal load is enormous ($\sim$ 400~W).  And another 1~W through the cooler neck-tube.  So we know that we will need to remove about 18~W with the shield cooler, but this is for a cold-tip working temperature of $\sim$115~K and the lift at this temperature is $\sim$~25~W, so the load is about 72~\% of the cooler capacity.

The detector and electronics receive radiation through the corrector and from the radiation shield, along with some internal heating.  The total is about 1.5~W, and is carried away, again, by the tripod legs.  The primary heat load carried by the tripod, however, is radiation falling on the tripod from the corrector and the thermal shield.  This is mitigated somewhat by the use of a low-thermal-emissivity black coating, NanoBlack, from Acktar\footnote {Acktar Advanced Coatings, www.acktar.com}.  The shield itself must be very black and will have high emissivity at thermal wavelengths.  This is mitigated by running the shield very cold, and the 'sweet spot' seems to be approximately 140~K.  At this shield temperature the radiation from the tripod to the shield is only 1~W, and the total load on the tripod is 2.5~W.  There is also conductive loss through the support tube, which is 2~W, and radiative loss of another 2~W, as well as about 1~W through the cooler neck-tube.  Hence, the total load on the detector cooler is about 7.5~W.  The cooler lift is $\sim$~15~W at an operating temperature of 80~K.  Therefore the cooler will run at approximately 50~\% capacity.

%%-----------------------------------------------------------
\subsection{Thermal background}
\label{Thermal_background} 

The thermal background is a very serious problem.  The requirement for the background flux is that it be less than 0.005~electrons/sec/pixel with an ambient temperature of 5~C.  To achieve this, the detector must see cold surfaces at all angles which do not actually contain the beam -- and those cold surfaces must extend all the way to the entrance aperture of the camera.  Hence we have a cooled radiation shield, a cold mirror, and the last corrector element floats cold as well.  The radiation shield is directly cooled by one of the two cryocoolers at the rear of the cryostat.  The shield will be set to a temperature of $\sim$ 140~K, well below the temperature needed to mitigate background (190~K would suffice).  The lower temperature helps cool the mirror (to 160~K) and reduces the radiative load on the detector tripod.

External background radiation comes from one of three sources.

%-------------
%-------------

First, radiation in the beam.  This radiation comes from the warm spectrograph and must be removed by filtration.  To accomplish this, a thermal rejection filter, a dichroic, is placed on the rear surface of the second corrector element.  The filter transmits in-band light and reflects unwanted light longward of 1.28~$\mu$m, rejecting at the level of a few parts per 10,000.  A second dichroic filter on the back of the Mangin mirror transmits unwanted wavelengths longward of 1.3~$\mu$m to a black beam dump behind the mirror.  Reflectivity for the Mangin coating wavelengths longer than 1.35~$\mu$m is of order 0.1~\%.  Data for the dichroic coatings described here are for coating designs by Infinite Optics, Inc.\footnote {Infinite Optics, Inc. www.infiniteoptics.com}

Second, radiation outside the beam that is {\it reflected} first by the radiation shield, then by the mirror, and ultimately incident on the detector.  This light enters the corrector from the outside world at large angles at which the coating does not reject so well.  Overall the background contribution from this source is small compared to the in-beam, in-band radiation.  To mitigate this radiation, baffles have been added to the inside of the radiation shield, which will be lined with Acktar Metal Velvet foil.

Third, radiation entering the cryostat at large angles and directly {\it scattered} by the radiation shield to the detector.  This source of background is only an issue for the rear cylindrical portion of the radiation shield, since the forward section cannot be viewed by the detector.  This source of background is small as well compared to the in-beam, in-band radiation, thanks largely to the very small diffuse reflectivity of the Metal Velvet lining.

\begin{figure}
   \begin{center}
   \begin{tabular}{c}
   \includegraphics[scale=0.45]{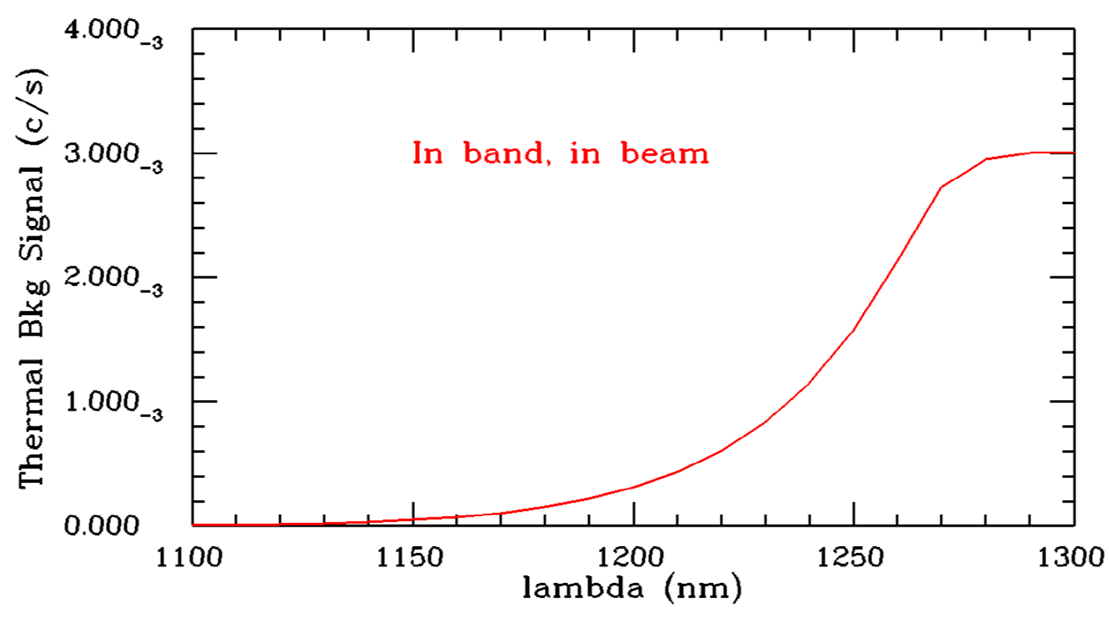}
   \end{tabular}
   \end{center}
   \caption[example] 
%>>>> use \label inside caption to get Fig. number with \ref{}
   { \label{fig:Fig13} 
Cumulative thermal background in the beam, and in-band for the rejection filter designs by Infinite Optics.  At a given wavelength, the curve gives the cumulative flux in electrons/sec/pixel for all wavelengths shorter than that wavelength.  Without the rejection filters on the second corrector element and the mirror, the background rises to an order of magnitude above the required limit by 1.30~$\mu$m, and it would rise to 30,000 times the required limit by the detector cutoff at 1.7~$\mu$m.}
   \end{figure}

Figure \ref{fig:Fig13} shows the cumulative thermal background in the beam, and in-band, with the rejection filter designs by Infinite Optics.  At a given wavelength, the curve gives the cumulative flux in electrons/sec/pixel for all wavelengths shorter than that wavelength.  Without the filters, the background would continue to rise rather than plateau at $\sim$~1.27~$\mu$m, rising to an order of magnitude above the required limit by 1.30~$\mu$m, and it would rise to 30,000 times the required limit by the detector cutoff at 1.7~$\mu$m.

%-------------
%-------------

%%-----------------------------------------------------------
\subsection{Cooldown and warmup}
\label{sec:Cooldown_and_warmup} 

The thermal dynamics are complex and to date we have not done detailed calculations to determine the cooldown time.  Rough calculations indicate that the detector will reach operating temperature in about 16 hours.  The mirror on the other hand is very massive and is only radiatively coupled and has a time constant of about 12 hours.  Hence to reach thermal equilibrium we estimate it will take approximately 48 hours.  Warming the camera will be achieved using 5~W heaters mounted on the beam dump and thermal pan.  It is estimated that it will take 70 hours to warm up.

To avoid condensing residual water on the detector during cooldown, the detector temperature will be biased warm at the start of the cooldown such that its temperature lags the radiation shield.  In a similar, we will start the warmup by heating the detector such that its temperature is warmer than the shield when the residual water vapor is released.

%%-----------------------------------------------------------
\subsection{Temperature sensors}
\label{Temperature_sensors} 
The thermal system implements 14 platinum RTDs (model HEL-705 from Honeywell).  These sensors are placed at strategic locations throughout the cryostat.  Of particular interest are the temperatures of the detector, ASIC, radiation shield, mirror cell, mirror, thermal pan, cryocooler cold tips, electronics enclosure, and the front bell of the cryostat, which provides an ambient reading.

%%%%%%%%%%%%%%%%%%%%%%%%%%%%%%%%%%%%%%%%%%%%%%%%%%%%%%%%%%%%%
\section{VACUUM SYSTEM}
\label{sec:VACUUM_SYSTEM} 

The vacuum system is relatively straightforward.  The cryostat volume is approximately 150 liters and the cryostat interfaces are sealed using baked Viton O-rings; the use of baked O-rings is critical to achieving the low permeation rates expected of Viton.  We will operate at a pressure of roughly 1~$\mu$Torr and monitor the pressure using a MPT200 Pirani/cold-cathode combination gauge from Pfeiffer mounted to the rear of the cryostat.  Vacuum is maintained by two Varian VacIon Plus 20 ion pumps mounted to the rear cover of the cryostat.  The lifetime of these ion pumps depends on the ultimate vacuum and is several years (80,000 hours) at one microtorr.  Given that thermal performance degrades significantly at poorer vacuum levels, one microtorr is a reasonable goal.  

The required pumping speed to maintain vacuum is dominated by permeation through the O-rings, of which there is roughly 13~m of total length.  Therefore given a permeation rate of 9.8 x 10-7~Torr-liters/sec/m\footnote{data taken from article by Phil Danielson, http://www.vacuumlab.com} and a base pressure of a microtorr, the required pumping speed is 12.8~liter/sec.  Hence, for the two pumps, the combined 40 liter/sec pumping speed provides, roughly, a 3x margin on the expected gas load from permeation.

To pump down from atomospheric pressure we will use a 60 liter/sec turbopump mounted to the rear of the cryostat.  The backing pump for the turbo will be located outside the spectrograph room.  A pneumatically controlled gate valve isolates the turbopump from the cryostat once a stable vacuum is reached and the cryostat is cooled to operating temperature.  Placing the turbopump directly on the rear of the cryostat maximizes the pumping speed, and reduces pump-down time.  Note that considerable pumping speed is often lost through low-conductance vacuum lines connecting pump-carts to cryostats.

In operation, we will pump to roughly 10~$\mu$Torr before cooling.  Pumpdown time depends on a number of factors but is clearly dominated by desorption of water from the various surfaces inside the cryostat.  In this design, as in many cryostat designs, the multi-layer insulation dominates the surface area and we estimate the total surface area to be roughly 100~m$^{2}$.  Given that, and the natural desorption rate of water, the amount of time required to pump from ambient to 10~$\mu$Torr is roughly 12 hours with a 60 liter/sec turbo.  Note that the turbo pumping speed is effectively reduced by the permeation rate through the O-rings; hence the effective pumping speed is only $\sim$~47~liters/sec with the 60~liter/sec pump.

%%%%%%%%%%%%%%%%%%%%%%%%%%%%%%%%%%%%%%%%%%%%%%%%%%%%%%%%%%%%%
\section{DETECTOR}
\label{DETECTOR} 

\begin{figure}
   \begin{center}
   \begin{tabular}{c}
   \includegraphics[scale=0.4]{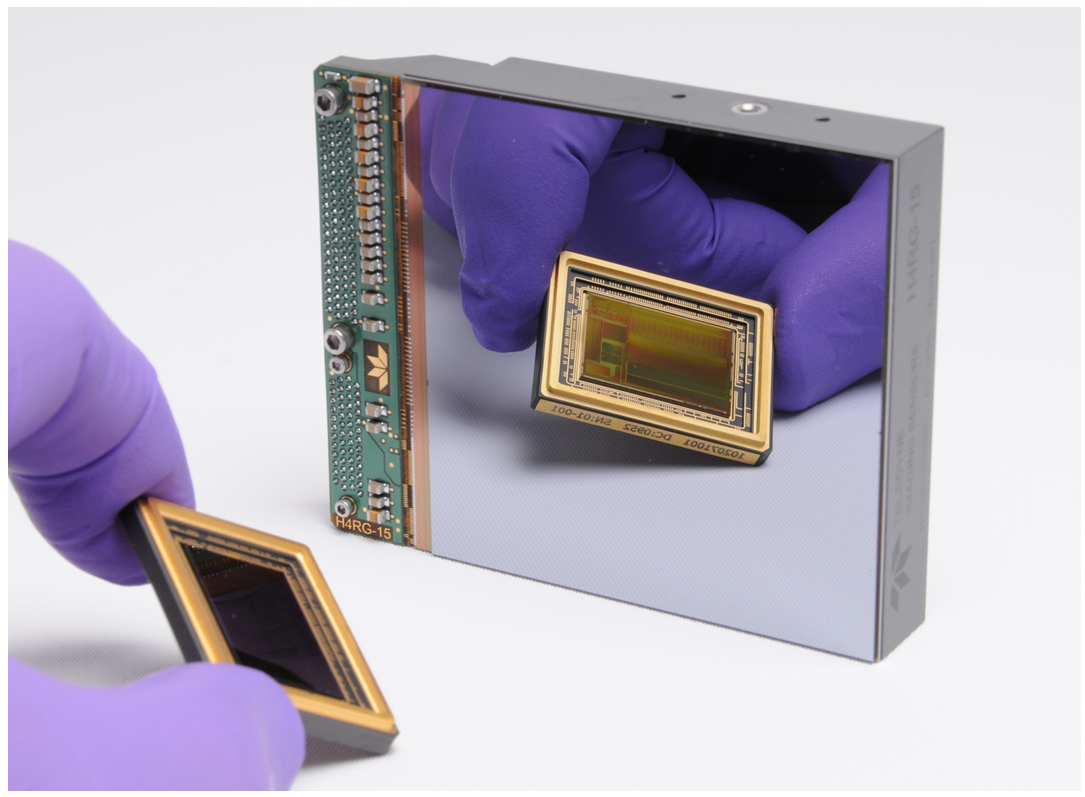}
   \end{tabular}
   \end{center}
   \caption[example] 
%>>>> use \label inside caption to get Fig. number with \ref{}
   { \label{fig:Fig14} 
Photograph of the Teledyne H4RG-15 showing the circuitry on one edge.  The gloved hand in the foreground is showing the new hermetic package for the sidecar signal processing ASIC, which we discuss below in the text.}
   \end{figure}

The IR detector chosen for the near infrared camera is the Teledyne Hawaii 4RG-15 HgCdTe device made with 1.7~$\mu$m cutoff material; there is really no alternative available, and the device appears to be a really superb match to our requirements.  The geometry of the active area of the device is essentially identical to the pair of Hamamatsu CCDs used in the two visible channel camreas: an array of 4096 x 4096 15~$\mu$m pixels.  The package is not square, but has an approximately 1~cm extension for connectors and bypass capacitors on one edge.  This costs us a bit of light (about 2~\%).  A photograph of the device is shown in Figure \ref{fig:Fig14}, along with the “Sidecar” signal-processing ASIC chip we will discuss in the electronics section.  The Sidecar is the only electronics module in the vacuum necessary for these detectors.

%-------------
%-------------

The H4RG has the following characteristics which are of interest to us:

%%-----------------------------------------------------------
\subsection{Quantum efficiency}
\label{sec:Quantum_efficiency}

\begin{figure}
   \begin{center}
   \begin{tabular}{c}
   \includegraphics[scale=0.4]{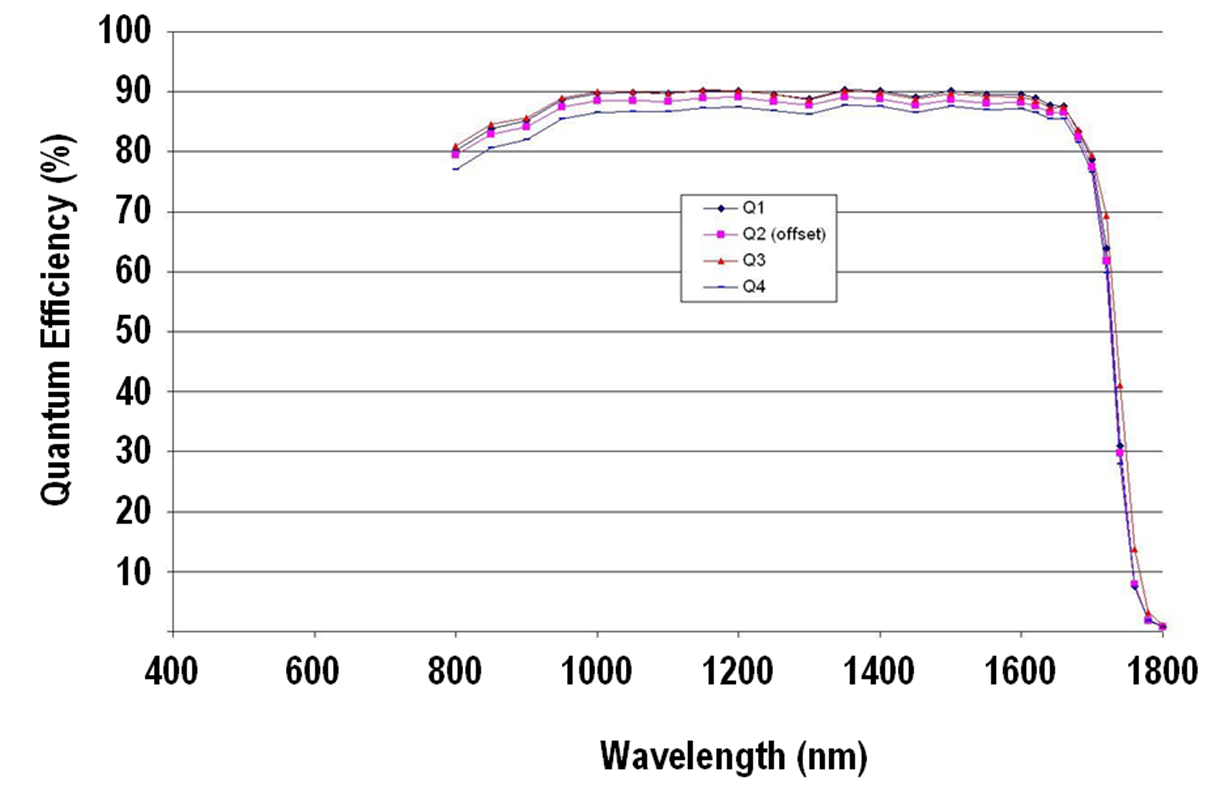}
   \end{tabular}
   \end{center}
   \caption[example] 
%>>>> use \label inside caption to get Fig. number with \ref{}
   { \label{fig:Fig15} 
The quantum efficiency of four 1.7~$\mu$m Hg series detectors, using the same substrate-removed technology as the H4RG-15 uses.  This data is from measurements taken at the NASA Goddard Space Flight Center and provided by Teledyne}
   \end{figure}

The measured QE is approximately 87~\%, very constant, over the wavelength range of interest for us.  A figure showing QE test results for 4 tested devices is given in Figure \ref{fig:Fig15} (measurements were taken by the NASA Goddard Space Flight Center and provided by Teledyne).  Of some interest to us is the quantum efficiency at longer wavelengths than the cutoff because of the very high thermal background photon flux.  This may be an item of some concern, but measurements made at ESO\cite{Finger2002} with similar devices show no measureable response beyond the cutoff.  This appears to be a desirable property of devices made with molecular beam epitaxy (MBE) as these devices are.  As discussed above, we need very strong filtration between our cutoff at 1.26~$\mu$m and the chip cutoff at 1.7~$\mu$m in order to have a satisfactory thermal background.

%-------------
%-------------

%%-----------------------------------------------------------
\subsection{Dark currrent}
\label{Dark_current} 
The measured dark current for 1.7~$\mu$m devices at 120~K is in the neighborhood of 0.005~electrons/sec/pixel.  This is to be compared with roughly 0.03~electrons/sec/pixel from the sky in the middle of the IR range.  We would like to reduce this somewhat and will operate at 100~K.

%%-----------------------------------------------------------
\subsection{Flatness}
\label{Flatness} 
The chips are specified to have a flatness of 20~$\mu$m peak-to-valley, but in fact production chips are all flatter than 12~$\mu$m, and Teledyne is willing to accept this as a no-cost spec option.

%%-----------------------------------------------------------
\subsection{Outputs}
\label{sec:Outputs} 
32 independent channels can be read at frequencies up to 400~kHz.  Since there are 16 million pixels, there are 500,000 pixels per channel.  At 100~kHz, then, a read takes 5 seconds.

%%-----------------------------------------------------------
\subsection{Read noise}
\label{sec:Read_noise} 
The 1.7~$\mu$m material is inherently noisier than the more common 2.5~$\mu$m material.  Teledyne has achieved 12 electron CDS (single-read) noise with the latter material, but the 1.7~$\mu$m material now yields noise of about 18 electrons for a single CDS read, and this is the figure we should plan on.

The technique to reduce this in IR devices is {\it Fowler sampling}.  Since the devices can be read non-destructively, one can read many times at the beginning and end of an exposure to establish charge levels.  It is common to do 32 reads at each end, and obtain a "Fowler-32'' result.  This should result in a factor of 4 lower noise than a single CDS read, and the measured result of 3 electrons for the 2.5~$\mu$m devices is what one expects.  For our devices, the expected number is 4.5 electrons, only slightly worse than the 4 electrons assumed.  Another read technique is called up-the ramp, in which the chip is read continuously while the exposure is proceeding.  This generates very large amounts of data but results in better noise performance than Fowler sampling (supposing one reads at the maximum rate in both cases) simply because one has more data, and in addition provides a mechanism for real-time rejection of cosmic ray events.  We will investigate both options.

\begin{figure}
   \begin{center}
   \begin{tabular}{c}
   \includegraphics[scale=0.45]{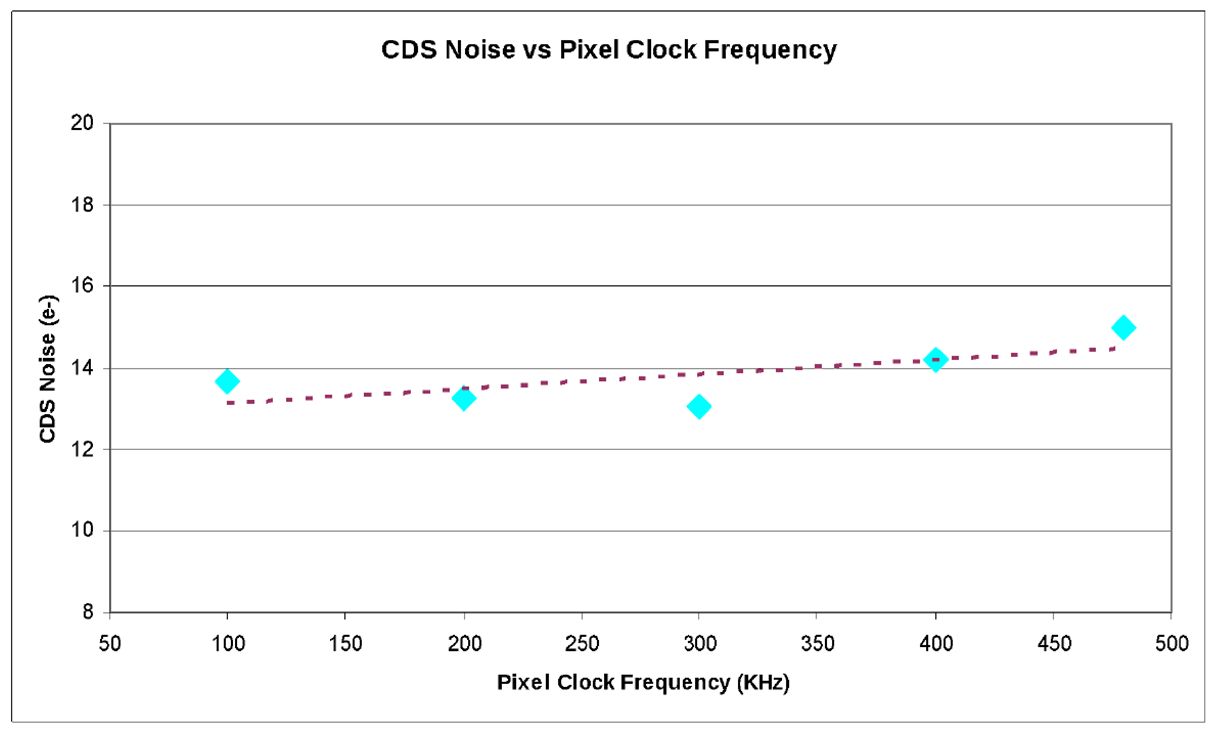}
   \end{tabular}
   \end{center}
   \caption[example] 
%>>>> use \label inside caption to get Fig. number with \ref{}
   { \label{fig:Fig16} 
Single-read CDS read noise for a 2.5~$\mu$m H2RG device using a Sidecar signal processor versus pixel rate.  Within the errors, there is no penalty for increasing the rate from the standard 100~kHz to 200 or even 300~kHz.  The resulting data rates are larger, of course, because of the 32 channels: About a 10~MHz total pixel rate at 300 kHz.  The data shown are for an engineering grade 2.5~$\mu$m cutoff (SWIR) H2RG SCA; 77 K, 32 outputs, slow mode, buffered output, 250~mV bias with Sidecar ASIC focal plane electronics.}
   \end{figure}

For Fowler sampling, 32 reads at each end takes about 2.5 minutes at each end, which is acceptable, but we can probably do better.  The read noise observed is very flat with read frequency, as discussed by Blank et al.\cite{2012SPIE.8453E..10B} and shown in Figure \ref{fig:Fig16}.  Therefore we should be able to read with essentially no penalty at 200~kHz, and do Fowler-64 sampling, which should reduce our noise to just above 3 electrons.

%-------------
%-------------

%%-----------------------------------------------------------
\subsection{Image persistence}
\label{sec:Image_persistence} 
Image persistence is a known problem with all HgCdTe devices.  Teledyne has been working on this, and the newest devices are much better than the older ones.  This is not really a significant problem with spectrographs in any case, but means the spectrograph {\it must} have a shutter, and must control the spectral samples so that excessively bright objects are not placed on the chip.  The severity of the effect must be evaluated during test.

%%%%%%%%%%%%%%%%%%%%%%%%%%%%%%%%%%%%%%%%%%%%%%%%%%%%%%%%%%%%%
\section{ELECTRONICS}
\label{sec:ELECTRONICS}

%%-----------------------------------------------------------
\subsection{Detector electronics} 
\label{sec:Detector_electronics}

We will use the Teledyne Sidecar ASIC Module (SMd) to read the H4RG detector.  This package and the board is just now becoming available.  It will use a relatively new column-grid-array (CGA) technology, which is similar to the standard ball-grid-array (BGA) mounting which is used for microprocessors and other products with very large numbers of contacts in a dense array, but the balls are replaced with relatively thin columns; the result is a system with greatly improved tolerance to the thermal mismatch between the ceramic package and the composite laminate circuit board.

The sidecar is mounted immediately behind the detector and will operate at the detector temperature.  It dissipates only about 150~mW at 100~kHz readout rate; that power scales very nearly linearly with the frequency.  In addition, it requires only 3.3~VDC at very small current to operate, and generates complete telemetry of voltages and operating state internally.  The sidecar and detector are connected by a short flex cable.  A multi-layer flex cable routes the signals from the sidecar to a hermetic feedthrough on the front ring of the cryostat.  A cable on the outside of the cryostat carries the signals to the Sidecar Acquisition Module (SAM), also a new Teledyne product, which generates timing, voltages, provides memory for image storage, and communicates to the outside world through a variety of protocols.  The SAM is mounted in a custom housing on the outside of the cryostat in close proximity to the feedthrough.

%%-----------------------------------------------------------
\subsection{The utility electronics}
\label{sec:The_utility_electronics} 
\begin{figure}
   \begin{center}
   \begin{tabular}{c}
   \includegraphics[scale=0.5]{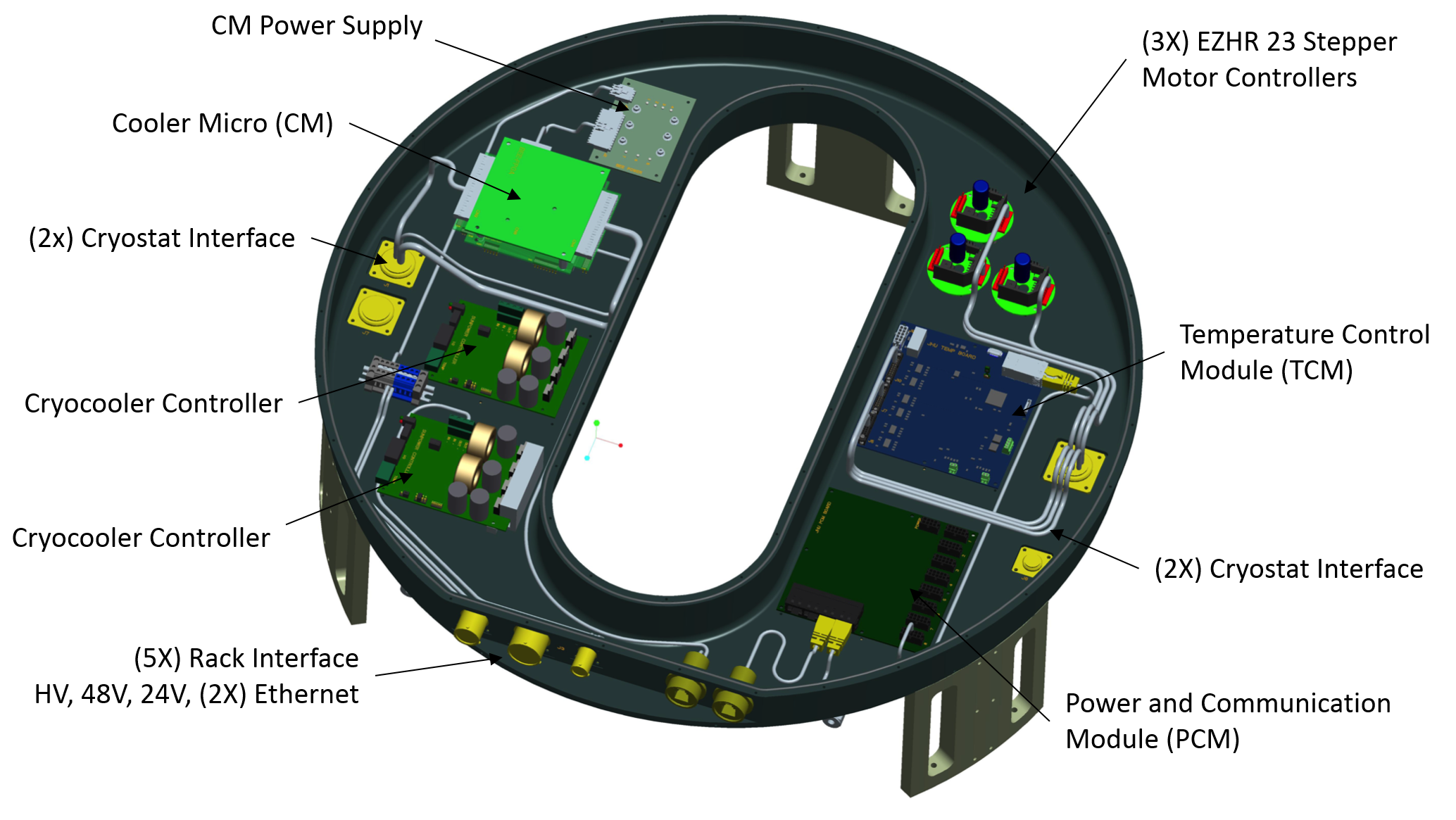}
   \end{tabular}
   \end{center}
   \caption[example] 
%>>>> use \label inside caption to get Fig. number with \ref{}
   { \label{fig:Fig17} 
Isometric view of the electronics enclosure for the near infrared camera with the cover removed.  The sealed enclosure contains control electronics for both cryocoolers, a temperature control module for readout of temperature sensors and control of heaters, a power control module for power distribution and Ethernet communication, stepper motor controllers for the focus mechanism, and a microprocessor for communication to the cryocooler controllers.  Heat generated by these components is removed by a glycol cooling loop affixed to the back side of the enclosure as seen in this view.}
   \end{figure}

Most of the electronics that monitor and control the general health of the camera are located in an electronics enclosure mounted to the rear cover of the cryostat; see Figure 2.  The custom enclosure sits behind the vacuum pumps.  It has a large oval relief in the center so as not to interfere with the cryocoolers, and to allow access to the turbopump roughing port.  Contained within the enclosure are the controllers for the two cryocoolers, a temperature control module for monitoring the RTDs and controlling heater power, a power control module to distribute power to the various devices 
and handle Ethernet communication, three motor controllers for the focus mechanism, and a microprocessor for controlling the coolers and its associated supply.  Power for all of these components comes from supplies in a utility rack outside the spectrograph room.  Combined, the devices inside the enclosure dissipate roughly 66 W.  This heat is removed by a glycol loop affixed to the cryostat-facing-side of the enclosure.  A rendered image of the enclosure is shown in Figure \ref{fig:Fig17}.

The high voltage supplies for the ion pumps are housed in a utility rack outside the spectrograph room.  Also in the utility rack are the 48~V and 24~V power supplies which feed power to the camera.  The 48~V supply feeds the cryocooler controllers.  The 24~V supply feeds the remaining components.

%-------------
%-------------

%%%%%%%%%%%%%%%%%%%%%%%%%%%%%%%%%%%%%%%%%%%%%%%%%%%%%%%%%%%%%
\section{SUMMARY}
\label{sec:SUMMARY} 

We have described the design of the near infrared camera for the PFS spectrograph, and have discussed the expected performance.  The spectrograph requires a fast camera that produces very good images over a wide field, with very little thermal background, and this has led to a number of engineering challenges.  We have presented a design that we believe satisfies all the variuos requirements without posing significant engineering difficulty.  Furthermore, we believe there are no thermal or electrical considerations which cause significant performance degradation beyond what the natural backgrounds and irreducible read noise in the detectors impose a priori.

%These are, in turn, consistent with the science requirements, if we blindly pretend that the requirements flow that direction, instead of the truth, which we all know, that %they flow the other way.

%%%%%%%%%%%%%%%%%%%%%%%%%%%%%%%%%%%%%%%%%%%%%%%%%%%%%%%%%%%%%
\acknowledgments     %>>>> equivalent to \section*{ACKNOWLEDGMENTS}       
 
We gratefully acknowledge support from the Funding Program for World-Leading Innovative R\&D in Science and Technology (FIRST), program:
"Subaru Measurements of Images and Redshifts (SuMIRe)", CSTP, Japan

%%%%%%%%%%%%%%%%%%%%%%%%%%%%%%%%%%%%%%%%%%%%%%%%%%%%%%%%%%%%%
%%%%% References %%%%%

\bibliography{report}   %>>>> bibliography data in report.bib
\bibliographystyle{spiebib}   %>>>> makes bibtex use spiebib.bst

\end{document}